\documentclass[useAMS,usenatbib]{mn2e}

\usepackage{graphicx}

\newcommand{\Msun}{$M_{\odot}$}
\newcommand{\Rsun}{$R_{\odot}$}

\newcommand{\kms}{km\,s$^{-1}$}
\newcommand{\vs}{$v \sin i$}
\newcommand{\teff}{$T_{\rm eff}$}
\newcommand{\lgg}{$\log\,{g}$}
\newcommand{\bz}{$B_{\ell}$}

\title[Orbit, composition, and magnetic field of HD 98088]
{Orbital parameters, chemical composition, and magnetic field of the Ap binary HD 98088\thanks{Based on observations obtained at the Bernard Lyot Telescope (TBL, Pic du Midi, France) of the Midi-Pyr\'en\'ees Observatory, which is operated by the Institut National des Sciences de l'Univers of the Centre National de la Recherche Scientifique of France. }
}

\author[Folsom et al.]{C.P. Folsom$^{1}$\thanks{E-mail: cpf@arm.ac.uk}, K. Likuski$^{2}$, G.A. Wade$^{2}$, 
O. Kochukhov$^{3}$,  E. Alecian$^{4}$, 
D. Shulyak$^{5}$\\
$^{1}$Armagh Observatory, College Hill, Armagh Northern Ireland BT61 9DG\\
$^{2}$Department of Physics, Royal Military College of Canada, P.O. Box 17000, Station `Forces', Kingston, Ontario, Canada, K7K 7B4\\
$^{3}$Department of Astronomy and Space Physics, Uppsala University, 751 20 Uppsala, Sweden \\
$^{4}$Observatoire de Paris, LESIA, 5, place Jules Janssen, F-92195 Meudon Principal CEDEX, France\\
$^{5}$Institute of Astrophysics, Georg-August-University, Friedrich-Hund-Platz 1, D-37077 G\"ottingen, Germany}

\begin{document}

\date{Received: 2012; Accepted: 2013}

\pagerange{\pageref{firstpage}--\pageref{lastpage}} \pubyear{2013}

\maketitle

\label{firstpage}

\begin{abstract}
HD 98088 is a synchronised, double-lined spectroscopic binary system with 
a magnetic Ap primary component and an Am secondary component.  
We study this rare system using high-resolution MuSiCoS spectropolarimetric data, 
to gain insight into the effect of binarity on the origin of stellar 
magnetism and the formation of 
chemical peculiarities in A-type stars. Using a new collection of 29 
high-resolution Stokes $VQU$ spectra we re-derive the orbital and 
stellar physical parameters and conduct the first disentangling of spectroscopic 
observations of the system to conduct spectral analysis of the individual 
stellar components. From this analysis we determine the projected rotational 
velocities of the stars and conduct a detailed chemical abundance analysis 
of each component using both the {\sc Synth3} and {\sc Zeeman} spectrum synthesis codes. 
The surface abundances of the primary component are typical of a cool Ap star, 
while those of the secondary component are typical of 
an Am star. We present the first magnetic analysis of both components using modern data. 
Using Least-Squares Deconvolution, we extract the longitudinal magnetic field 
strength of the primary component, which is observed to vary between +1170 and -920 G 
with a period consistent with the orbital period. There is no field detected in the
secondary component.  The magnetic field in the primary is predominantly dipolar, 
with the positive pole oriented approximately towards the secondary.  
\end{abstract}


\section{Introduction}
Ap stars are magnetic intermediate-mass main sequence stars that exhibit distinctive 
chemical peculiarities in their atmospheres. Although other classes of chemically 
peculiar stars exist (e.g. Am stars, Hg-Mn stars), these stars have 
been demonstrated to lack strong, organised magnetic fields at their surfaces 
\citep[e.g.][]{Shorlin2002,Wade2006-alphaAnd,Makaganiuk2011-HgMn-magnetic-survey}.  
Ap stars appear to be the only class of middle main-sequence stars for which, 
in all cases, an observable magnetic field is present \citep{Auriere2007}. 

The magnetic fields of Ap stars have been established to have important global 
dipole components ranging in strength from hundreds to tens of thousands of gauss. 
The symmetry axis of the dipole component is almost always tilted relative the 
stellar rotation axis. In addition, Ap stars generally spin much more slowly than 
non-peculiar stars of similar masses \citep{Abt1995-Astar-vsini,stepien2000-Ap-rotation-momentum}, 
and as they spin they exhibit line profile variations attributed to rotational modulation of patchy, 
non-axisymmetric lateral and vertical distributions of chemical abundance in their photospheres. 
While it is generally accepted that the fundamental mechanism 
responsible for the chemical peculiarities is microscopic chemical diffusion 
(as described by \citealt{Michaud1970-diffusion}), the origin of chemical patchiness, 
and the relationship to the magnetic field, is poorly understood. 

Although modern surveys \citep{Gerbaldi1985-Ap-binary-frequency,Carrier2002-binarity_in_Ap} 
indicate that Ap stars exhibit about the same binary frequency as normal A-type stars 
\citep[$47\pm5$\%,][]{Jaschek1970-old-binary-frequency}, they generally agree that 
there is a clear lack of Ap stars in short-period (i.e. $P_{\rm orb}=3.0$ days or less) systems 
\citep{Carrier2002-binarity_in_Ap}. One might hypothesize that this could be because 
these systems tend to be synchronized.  Such a short-period binary would imply a 
short rotational period for both components.  Rapid rotation would induce rotational 
mixing, inhibiting the development of chemical peculiarities.  However, no similar 
absence of short period systems exists for Am binaries.  This suggests that binaries 
with short periods are capable of hosting chemically peculiar stars, but contain few 
magnetic (chemically peculiar) stars.  The few magnetic Ap stars reported to be in 
short period binaries could therefore be very interesting targets for studying the 
origin and evolution of magnetic fields in intermediate-mass stars.  

Systems containing two A-type stars are of particular interest, since they provide 
an opportunity for studying not only the incidence and evolution of stellar magnetic 
fields, but also of chemical peculiarity in both stellar components. 
A comprehensive literature review reveals that these ideal systems are indeed 
very rare.  Both HD~135728 \citep{Freyhammer2008-new-Ap+1binary} 
and HD~59435 \citep{Wade1999-hd59435-Ap-binary} have magnetic Ap secondary components; 
however their primary components are giants of spectral type G8.  While these primary giant 
G8 stars evolved from main sequence stars with masses similar to that of the Ap star, 
the evolved primary component makes these systems less useful to study.  There are only 
five reported SB2 binaries with a magnetic Ap primary component and a main sequence 
A star companion.  The first of these systems is HD~55719, which was studied by 
\citet{Bonsack1976-hd55719-Ap-binary}, although as a southern object it has received 
relatively less attention in the modern literature.  
HD~5550, HD~22128, and HD~56495  \citep[e.g.][]{Carrier2002-binarity_in_Ap} 
show composite A-type spectra, although the magnetic nature of the 
Ap component has yet to be definitively established.  
The final system is HD~98088, which is the subject of this paper.  
This system is very well-suited for the study of magnetic fields and chemical 
peculiarities, and is one of the brightest and best-studied magnetic binary systems.

HD 98088 (HR 4369, HIP 55106) was first observed as a spectroscopic binary by 
\citet{Abt1953-hd98088-vr}, although he reported it to be an SB1.  The primary radial velocity 
was found to vary with an orbital period of $5.905\pm0.005$ days.  
Spectrophotometric analysis using the Mills spectrograph 
showed variability in beryllium and titanium lines.  Five years later 
\citet{Babcock1958-magnetic-catalog} confirmed the presence of a magnetic field in HD~98088 
using polarized spectroscopy.  The strength of the longitudinal magnetic field was found to vary 
between $-1000$ and $+800$ gauss with a period consistent with the orbital period 
derived by \citet{Abt1953-hd98088-vr}.  These measurements made HD~98088 the first observed 
synchronized spectroscopic binary containing a magnetic star \citep{Babcock1958-magnetic-catalog}.  
In his \citeyear{Osawa1965-Ap-classification} paper on spectral classification, 
\citeauthor{Osawa1965-Ap-classification} classified the Ap 
primary star to be of spectral subtype SrCr.   

The first complete orbital solution of this SB2 system was computed by \citet{Abt1968-hd98088-vr-orbit}. 
\citet{Abt1968-hd98088-vr-orbit} searched for photometric eclipses without success.  
From the the difference in equivalent widths of the Ca {\sc ii} K, Mg {\sc ii} ${\lambda}4481$, 
Na {\sc i} D$_{1}$ and D$_{2}$, Si {\sc ii} ${\lambda}6371$, and H${\alpha}$ lines, 
the magnitude difference of the components were estimated to be 
${\Delta}m_{V}=1.2$.  This led them to conclude that the primary has spectral type 
A3Vp and the secondary has spectral type A8V.  
The masses of each component were then inferred from the spectral type, 
and combined with the dynamic $M{\sin}^{3}i$ to determine the system's inclination angle $i=67\degr$. 
An interesting feature discovered by \citet{Abt1968-hd98088-vr-orbit} 
is that despite its short orbital period,  HD~98088 has significant orbital eccentricity.  
This suggests that apsidal motion should take place. 
However, \citet{Wolff1974-hd98088-vr} obtained new 
spectroscopic observations that were used in conjunction with older measurements to refine 
the period of apsidal motion to at least 700 years. 

\citet{Carrier2002-binarity_in_Ap} reobserved the system using the CORAVEL 
velocimeter.  With these measurements, and all other published radial velocity 
measurements of the system, the orbital period of HD 98088 was refined to 
$5.905111\pm0.000004$ days.  The \cite{Carrier2002-binarity_in_Ap} solution is 
consistent with the solution of \citet{Abt1968-hd98088-vr-orbit}.   \citet{Carrier2002-binarity_in_Ap} 
observed no significant variation of the longitude of periastron, and concluded that 
the apsidal period is longer than one millennium.  
They used the Hipparcos distance and effective temperatures determined from Geneva photometry
to derived the masses of the components from their HR diagram 
positions to be $M_{1}=2.261\pm0.093$ \Msun\ and $M_{2}=1.677\pm0.085$ \Msun. 
Assuming negligible Zeeman broadening and a synchronized system, they found the radii to 
be $R_{1}=3.27\pm0.43$ \Rsun\ and $R_{2}=2.10\pm0.28$ \Rsun\ using the 
projected rotational velocities and the orbital period.  The inclination angle was 
found to be $i=66^{\circ}$ by comparing the minimum dynamical masses to the 
evolutionary masses.  

The HD 98088 system potentially provides a very interesting and rare insight 
into chemical peculiarities and magnetism in A type stars, in the context 
of close binary systems.  The similarities between the physical 
parameters of the primary and secondary make it an excellent laboratory for 
comparative astrophysics.  In this paper we use spectroscopic and polarimetric data 
to improve the orbital parameters of the system, to perform a detailed abundances 
analysis of the spectra of both stellar components, and to analyze the magnetic 
properties of the system.

\section{Observations}
\label{observations}

Observations of HD 98088 were obtained with the Multi-Site Continuous Spectroscopy (MuSiCoS) 
spectropolarimeter, a now-decommissioned high resolution spectropolarimeter 
located at the the T\'elescope Bernard Lyot at the Observatoire du Pic du Midi, 
France.  The instrument consists of a bench mounted cross-dispersed \'echelle 
spectrograph \citep{Baudrand1992-musicos-spectrograph}, fibre-fed from a Cassegrain 
mounted polarimeter unit \citep{Donati1999-musicos-pol}.  
The instrument provides continuous wavelength coverage 
from about 4500 to 6500 \AA\, at a resolution of $R=35\,000$.  
Observations were obtained in spectropolarimetric mode, providing Stokes 
$V$, $Q$ and $U$ spectra as well as Stokes $I$ spectra.  The data were 
reduced using the ESpRIT \citep{Donati1997-major} reduction tool, 
which performs relevant calibrations and optimal 1D spectrum extraction.  

A total of 29 spectra -  10 Stokes $V$, 10 Stokes $Q$ and 9 Stokes $U$ - 
were obtained between Jan 1999 and Jan 2002.  
A summary of the observations is presented in Table~\ref{observations_table}.  
The typical signal-to-noise ratio (S/N) per 4.5 km/s spectral pixel was 
about 170 for Stokes $V$ and 230 for Stokes $Q/U$. The orbital 
phases are calculated with the ephemeris described in Table \ref{orbital-param}. 

Least-Squares Deconvolution (LSD) was applied to each of the calibrated and 
normalised spectra. Line masks were constructed using the derived atmospheric 
parameters (determined in Sect.\ \ref{params}) and abundances 
(derived in Sect.\ \ref{Chemical Abundances}) 
of both the primary and secondary components. Due to the similarity between 
the temperatures and abundances of the two stars, only small differences were 
observed in the profiles obtained from the two lines masks. Ultimately, 
we decided to use only the line mask corresponding to the primary to extract 
the LSD profiles. 

All LSD profiles are normalised to a wavelength of 525 nm and a land\'e 
factor of 1.41, corresponding approximately to the S/N-weighted mean of 
the observed lines included in the mask.  

\begin{table*}
\centering
\caption[Summary of MuSiCoS observations]
{Summary of MuSiCoS observations. The UT date, the Julian Date, the Stokes parameter, 
and the peak signal to noise ratio are given.  The orbital phase of the 
observation was calculated using the ephemeris described in Table \ref{orbital-param}.  
Since the system appears to be synchronized, the rotational phase is assumed 
to be equal to the orbital phase.  For each observation, four subexposures of $300$s 
duration were acquired.}
\begin{tabular}{ccccrrrrrr}
\hline\hline
Date & HJD            & Phase  & Stokes& Peak&  \multicolumn{2}{c}{Velocity (\kms)} \\
     & (2,450,000+)   &        &Param  & S/N &   Primary        & Secondary \\ 
\hline
Jan 18, 1999 &  1197.617 & 0.358 & $V$ & 190 & -8.0 $\pm$ 2.2   & \\                   
Jan 18, 1999 &  1197.643 & 0.362 & $Q$ & 240 & -9.2 $\pm$ 2.3   & \\
Jan 18, 1999 &  1197.674 & 0.368 & $U$ & 230 & -6.3 $\pm$ 2.2   & \\                   
Jan 23, 1999 &  1202.590 & 0.200 & $V$ & 140 & 52.5 $\pm$ 2.6   & -84.7 $\pm$ 1.5 \\   
Jan 23, 1999 &  1202.617 & 0.205 & $Q$ & 210 & 53.7 $\pm$ 2.7   & -85.9 $\pm$ 1.5 \\   
Jan 24, 1999 &  1203.555 & 0.364 & $V$ & 180 & -8.9 $\pm$ 2.2   & \\                   
Jan 24, 1999 &  1203.581 & 0.368 & $Q$ & 200 & -9.9 $\pm$ 2.3   & \\                   
Jan 24, 1999 &  1203.613 & 0.374 & $U$ & 200 & -7.7 $\pm$ 2.2   & \\                   
Mar 05, 2000 &  1609.524 & 0.112 & $V$ & 190 & 77.5 $\pm$ 2.2   & -118.7 $\pm$ 0.9 \\  
Mar 07, 2000 &  1611.519 & 0.450 & $Q$ & 230 & -33.7 $\pm$ 2.2  & 34.1 $\pm$ 0.5 \\    
Mar 07, 2000 &  1611.554 & 0.456 & $U$ & 240 & -35.3 $\pm$ 2.3  & 34.5 $\pm$ 0.5 \\    
Dec 05, 2001 &  2249.685 & 0.520 & $V$ & 100 & -53.2 $\pm$ 2.4  & \\                   
Dec 05, 2001 &  2249.710 & 0.525 & $Q$ & 160 & -54.0 $\pm$ 2.4  & \\                   
Dec 05, 2001 &  2249.742 & 0.530 & $U$ & 130 & -52.5 $\pm$ 2.4  & \\                   
Dec 07, 2001 &  2251.693 & 0.860 & $Q$ & 270 & -28.6 $\pm$ 2.3  & 29.1 $\pm$ 1.6 \\    
Dec 07, 2001 &  2251.725 & 0.866 & $U$ & 250 & -25.5 $\pm$ 2.4  & \\                   
Dec 07, 2001 &  2251.751 & 0.870 & $V$ & 180 & -23.3 $\pm$ 2.5  & \\                   
Dec 08, 2001 &  2252.690 & 0.029 & $Q$ & 230 & 65.8 $\pm$  2.4  & -101.7 $\pm$ 1.8 \\  
Dec 08, 2001 &  2252.722 & 0.035 & $U$ & 230 & 67.3 $\pm$  2.2  & -103.8 $\pm$ 2.5 \\  
Dec 08, 2001 &  2252.748 & 0.039 & $V$ & 160 & 68.8 $\pm$  2.5  & -107.0 $\pm$ 1.4 \\  
Dec 09, 2001 &  2253.687 & 0.198 & $Q$ & 250 & 54.3 $\pm$  2.8  & -86.0 $\pm$ 2.6 \\   
Dec 09, 2001 &  2253.715 & 0.203 & $U$ & 240 & 52.9 $\pm$  2.5  & -85.0 $\pm$ 1.8 \\   
Dec 09, 2001 &  2253.750 & 0.209 & $V$ & 230 & 50.6 $\pm$  2.8  & -84.4 $\pm$ 1.1 \\   
Dec 12, 2001 &  2256.710 & 0.710 & $Q$ & 270 & -71.5 $\pm$ 2.6  & 83.0 $\pm$ 5.6 \\    
Dec 12, 2001 &  2256.741 & 0.715 & $U$ & 250 & -71.3 $\pm$ 2.6  & 82.5 $\pm$ 2.8 \\    
Dec 12, 2001 &  2256.762 & 0.719 & $V$ & 110 & -71.2 $\pm$ 2.6  & 82.0 $\pm$ 1.8 \\    
Jan 05, 2002 &  2280.694 & 0.772 & $Q$ & 230 & -63.2 $\pm$ 2.5  & \\
Jan 05, 2002 &  2280.731 & 0.778 & $U$ & 200 & -61.3 $\pm$ 2.4  & \\
Jan 05, 2002 &  2280.759 & 0.783 & $V$ & 130 & -60.5 $\pm$ 2.4  & \\
\hline\hline
\end{tabular}
\label{observations_table}
\end{table*}

\section{Orbit}

Orbital parameters for HD~98088 were determined using new and previously 
published measurements of the radial velocities of 
both the primary and secondary components. Published data are from 
\citet{Abt1953-hd98088-vr}, \citet{Abt1968-hd98088-vr-orbit}, \citet{Wolff1974-hd98088-vr}, 
and \citet{Carrier2002-binarity_in_Ap}.  The new radial velocities 
measurements were obtained using the LSD profiles discussed in the 
previous section, extracted from each of the reduced spectra. 
To measure the radial velocities of the LSD line profiles of each component 
we used an IDL code, described by \citet{Alecian2008-HD200775}, that
fits the binary profile by least-squares using the IDL function CURVEFIT. 
The binary profile is assumed to be the sum of two photospheric functions, 
one for each component of the system. 
Each photospheric function is the convolution of a rotation profile and a Gaussian 
of instrumental and turbulent velocity width \citep[][]{Gray2005-Photospheres}. 
For the purpose of the fit, the turbulent velocity was fixed at 2 \kms, 
and the rotation profile broadening was left as a free parameter. 
The errors were calculated using a Monte-Carlo-type approach in which 
five thousand observed profiles were calculated by adding random noise to the 
best fit profile. The fitting procedure was repeated on each of these five thousand 
profiles and the standard deviation of these measurements was adopted as the error. 
The derived radial velocities are reported in Table \ref{observations_table}. 

\begin{figure*}
\centering
\includegraphics[width=4.0in,angle=90]{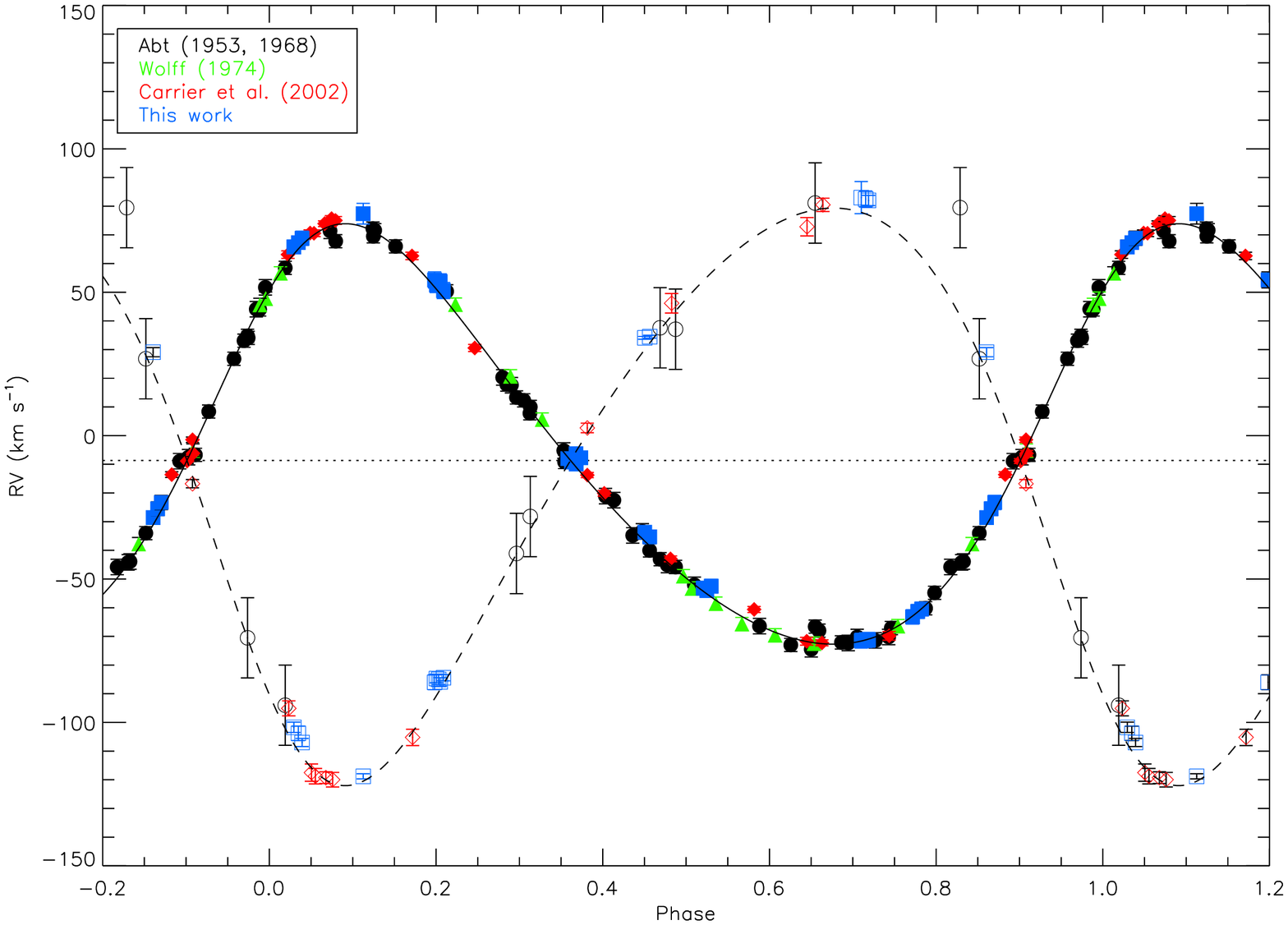}
\caption{Measured radial velocities for the primary (full symbols) and secondary (empty symbols), 
phased with the orbital period of $5.9051102 \pm 0.0000023$ days. 
Also shown is the best fit orbital curve to those velocities for the primary 
(solid line) and secondary (dashed line), using the ephemeris from Table \ref{orbital-param}.  }
\label{orbit-fit}
\end{figure*}

The full set of radial velocity measurements was used to determine orbital parameters 
for the binary system.  This was done by fitting orbital radial velocity curves 
to the measured velocities.  A small constant offset between our values 
and \citet{Carrier2002-binarity_in_Ap} was used ($\approx2$ \kms), 
significantly improving our fit to the radial velocities, 
following the work of \citet{Carrier2002-binarity_in_Ap} who found it necessary 
to shift their radial velocities with respect to previous measurements.
The best fit orbital parameters are presented in Table \ref{orbital-param}, and 
the fit to the radial velocities is shown in Fig.\ \ref{orbit-fit}. 
Our orbital parameters are consistent with those of \citet{Carrier2002-binarity_in_Ap}, 
although our parameters are more precise due to our larger dataset.

\begin{table}
\centering
\caption{Best fit orbital parameters, allowing for offsets between 
the radial velocities of \citet{Carrier2002-binarity_in_Ap} and older measurements ($\Delta v_1$), 
and between our measurements and older values ($\Delta v_2$). }
\label{orbital-param}
\begin{tabular}{ll}
\hline \hline \noalign{\smallskip}
Reduced $\chi^2$     & 1.2039\\
RMS(A)               & 2.1863\\
RMS(B)               & 6.9670\\
$P$ (d)              & $5.9051102 \pm 0.0000023$ \\
$T_0$ (HJD -2,400,000)& $34401.369 \pm 0.015$\\  
$K_{\rm A}$ (\kms)     & $73.31 \pm 0.26$\\ 
$K_{\rm B}$ (\kms)     & $100.70 \pm 0.43$\\
$V_0$ (\kms)         & $-8.65 \pm 0.29$\\
$e$                  & $0.1840 \pm 0.0025$\\
$\omega$ ($\degr$)   & $313.25 \pm 0.89$\\
$\Delta v_1$ (\kms) & $-1.34 \pm 0.38$\\
$\Delta v_2$ (\kms) & $-3.02 \pm 0.40$\\
$M_{\rm A}/M_{\rm B}$    & $1.374 \pm 0.008$\\
$M_{\rm A} \sin^3 i$ (\Msun)& $1.772 \pm 0.013$\\
$M_{\rm B} \sin^3 i$ (\Msun)& $1.2897 \pm 0.0089$\\
$a_{\rm A} \sin i$ (\Rsun)  & $8.411 \pm 0.030$\\
$a_{\rm B} \sin i$ (\Rsun)  & $11.553 \pm 0.050$\\
\noalign{\smallskip} \hline \hline
\end{tabular}
\end{table}

\section{Spectral Disentangling}

The spectra of HD~98088 exhibit a complex pattern of two similar absorption 
line systems, with velocity separation changing
from zero to $\sim$200~\kms\ and a period on the time scale of 5.9 days. In addition, 
one of the components has significant intrinsic line profile
variability. In this situation a spectral disentangling procedure is essential 
to separate this effect from variable line blending due
to orbital motion of the binary components and to obtain high-quality average 
spectra for abundance analysis.

We employ the direct spectral decomposition technique discussed by 
\cite{Folsom2010-AR-Aur}. Our algorithm operates in an iterative fashion, as follows. 
We begin with a set of approximate radial velocities for each component and initial guesses of 
their spectra.  Then the contribution of the less luminous component B is subtracted 
from the observed spectra, and all spectra are shifted to the rest frame of component A. 
The spectra are interpolated on a standard wavelength grid and co-added, 
yielding a new approximation of the primary spectrum.  
The same procedure is then used to update the approximation of the secondary spectrum.  
This sequence of operations is repeated up to convergence. 
Once converged spectra for the two components have been found 
the second part of the algorithm begins.  In this part we use a least-squares minimisation 
routine to derive improved radial velocities (if desired) and correct the 
continuum normalisation of the individual spectra using low-degree polynomials 
(up to this point the program does not distinguish between lines and continuum). 
The first part of the program calculating disentangled spectra and the second part of the program 
fitting radial velocities are alternated until the changes in parameters from one iteration 
to the next are below given thresholds.

This spectral disentangling was applied to overlapping 45~\AA\ segments of the spectra 
from 4500 \AA\ to 6590 \AA.  While this wavelength range includes the H$\alpha$ and H$\beta$ Balmer 
lines, disentangling does not produce useful results for these lines.  This is because the radial velocity 
variation between the components of HD 98088 is insufficient to disentangle such broad lines, 
and because automatic continuum normalisation of these broad features can be inaccurate.  
The radial velocities used were measured from the LSD profiles, presented in 
in Table~\ref{observations_table}, and were not refined in the disentangling process 
due to intrinsic variability of the line profiles in the primary.  

Disentangling yields separate averaged high-quality ($S/N\approx 1000$) spectra of components A and B, 
as well as standard deviation curves which characterise the remaining 
discrepancy between the observations and composite model spectra.  
The standard deviation can be examined in the rest frame of the primary and secondary separately. 
A coherent excess in the standard deviation corresponding to a certain absorption feature provides 
a straightforward, objective identification of intrinsic variability in that line. 
Performing this comparison, we found that all cases of intrinsic spectrum variability 
were associated with the lines of HD~98088~A.  
Variability in the primary was found in lines of Cr, Ba, Nd, and Eu, with weaker variability 
in lines of Mn and Sc.  No clear variability was found in other lines, at the level of the noise in our spectra.  

\section{Fundamental Parameters}
\label{params}

The luminosity ratio of HD~98088 was estimated using the H$\alpha$ Balmer line cores.  
While the Balmer lines wings of the two components of HD~98088 are heavily blended, 
the cores are clearly separated in most of our observations.  
The depths of the H$\alpha$ cores are only weakly dependent on temperature, 
and thus can be used to estimate the flux ratio of HD~98088~A to HD~98088~B at 6560 \AA.  

\begin{figure}
\includegraphics[width=3.3in]{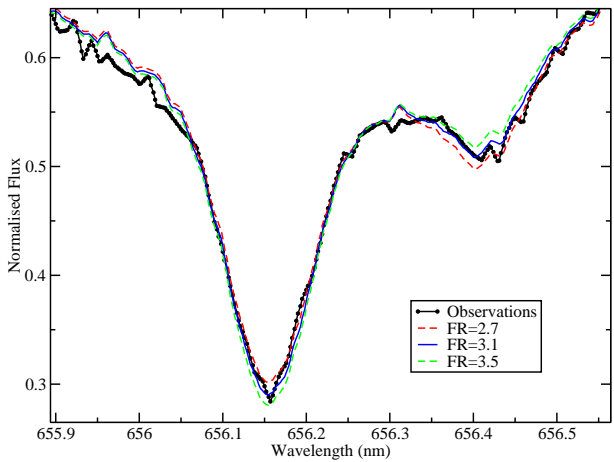}
\caption{Sample fits to the observed combined H$\alpha$ core of HD 98088, for our best fit 
flux ratio at 6560 \AA\ (FR) of 3.1, and for two other flux ratios at $\pm2\sigma$. }
\label{Halpha-fit}
\end{figure}

To model the H$\alpha$ line cores, we began with the \teff\ and \lgg\ estimates from 
\citet{Carrier2002-binarity_in_Ap}.  Rather than using synthetic H$\alpha$ profiles, 
which can have inaccurate cores due to significant non-LTE effects, we used observed H$\alpha$ 
profiles of the single stars HD 108651 and HD 73709.  The stars were observed 
by \citet{Shorlin2002} with MuSiCoS and have similar parameters to 
HD~98088~A (HD~108651: \teff~$=8100$~K, \lgg~$=4.36$, \vs~$=23$~\kms; HD~73709: \teff~$=8080$~K, 
\lgg~$=4.02$, \vs~$=22$~\kms; \citealt{Shorlin2002}).  Observations of the two stars were averaged 
to produce our model H$\alpha$ line core.  A composite binary H$\alpha$ line was 
constructed by adding two model H$\alpha$ lines, appropriately Doppler shifted, 
weighted by the flux ratio at 6560 \AA.  This was then fit to the observed H$\alpha$ line 
in the binary spectrum by varying the flux ratio.  This was repeated for seven observations, 
producing values of $F_A/F_B$ of 3.0 to 3.3.  An uncertainty was estimated by the 
range of values allowed by the noise in our observations, and our final 
best value is $F_A/F_B = 3.1 \pm 0.2$.  An example of a fit to the 
H$\alpha$ line core of an observation is given in Fig. \ref{Halpha-fit}.  

The flux ratio at 6560 \AA\ was then converted into a (bolometric) luminosity ratio 
by comparing with synthetic ATLAS9 flux distributions \citep{Kurucz1993-ATLAS9etc} 
for the atmospheric parameters in Table \ref{fundimental-param} at 6560 \AA, 
and assuming $L_{\rm A}/L_{\rm B} = (T_{\rm eff, A}/T_{\rm eff, B})^{4} (R_{\rm A}/R_{\rm B})^{2}$.  
This was done iteratively with the determination of effective temperature 
and surface gravity, discussed below.  Using our final \teff, 
8300 K for the primary and 7500 K for the secondary, this produces 
$L_{\rm A}/L_{\rm B} = 3.62 \pm 0.72$ and $R_{\rm A}/R_{\rm B} = 1.55 \pm 0.05$.  

Effective temperatures for both components of HD~98088 were determined from 
the spectral energy distribution (SED) of the system, since the Balmer lines were heavily blended. 
Since the system is not spatially resolved, simple photometry of the system 
is insufficient to determine the temperatures of both stars. 

Optical spectrophotometry was taken from the database of \citet{Adelman1989-spectrophot-cat}. 
Photometry in the UV was based on observations from the TD1 satellite, 
as presented by \citet{Thompson1978-TD1-fluxes}.  IR photometry was taken 
from the Two Micron All Sky Survey (2MASS) \citep{Skrutskie2006-2MASS}. 
Synthetic model fluxes were taken from ATLAS9 models.  Enhanced abundances 
($10 \times$ solar) were used to reflect the chemically peculiar nature of the primary 
and, as we will show, the secondary.  Both the models and the observations 
were normalised to the flux at 5000 \AA, and are presented in magnitudes. 
The model fluxes were added, weighted by the luminosity ratio. 

Since HD~98088 is nearby ($129.5\pm6.7$ pc) reddening of system is likely very small, 
and has thus been neglected in our analysis.  The reddening maps of \citet{Lucke1978-reddening-map} 
suggests that $E(B-V) < 0.02$.  The detailed maps of \citet{Schlegel1998-dust-map}, 
which include all Galactic reddening along a line of sight, 
puts an upper limit on the reddening of $E(B-V) < 0.05$, though the true value for the HD~98088 
system is likely much less.  Thus, following the recommendation of \citet{Lipski2008-ap-sed-fitting}, 
since the reddening is small we neglect it.  

\begin{figure*}
\includegraphics[width=5.5in]{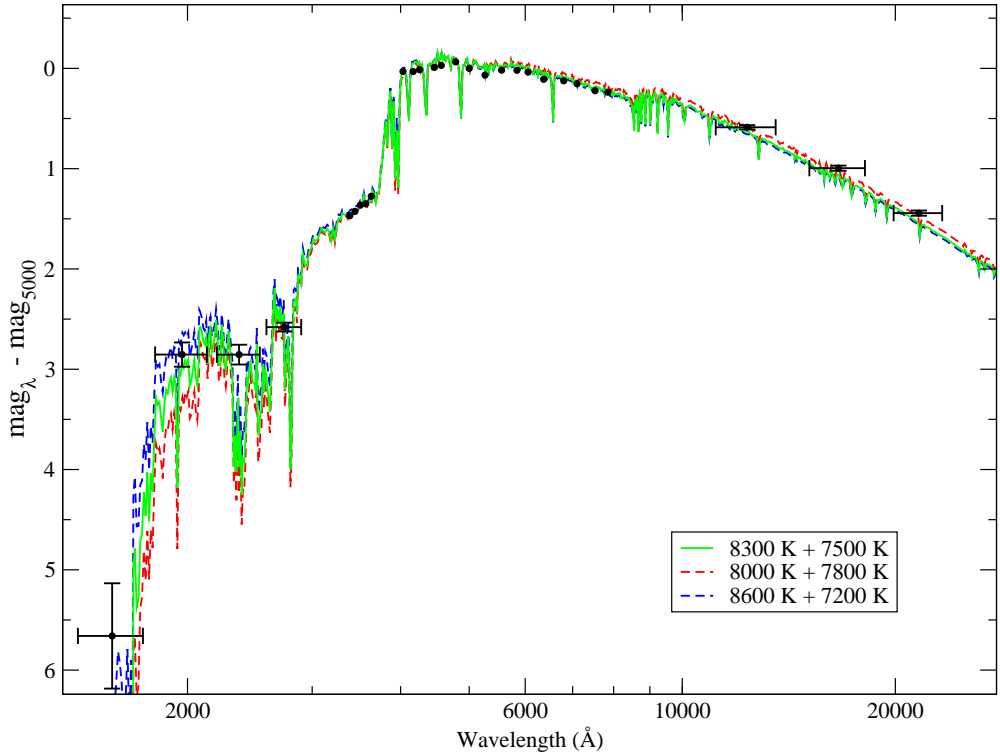}
\caption{Fits to the spectrophotometric data using model binary flux distributions.  
The models are labelled by the \teff\ of the primary and secondary, all for \lgg $=4.0$ 
and a luminosity ratio of 3.6.  The error bars in wavelength represent the full width of 
the filters for the UV and IR photometry. }
\label{sed-fit}
\end{figure*}

Fitting the SED provided well constrained effective temperatures for both stars.  
Since the primary dominates the flux in the UV, and the secondary's flux nearly matches the primary 
in the IR, precise \teff\ values for both stars can be derived.  However, surface gravity 
cannot be determined simultaneously for both stars with the same precision, since \lgg\ 
predominately affects the Balmer jump where flux from both stars is important.  
The luminosity weighted mean \lgg\ is 4.0, but there is a mild degeneracy between the two stars. 
The fit to the observed SED is presented in Fig. \ref{sed-fit}, and the best fit 
parameters for the primary are \teff\ $= 8300 \pm 300$ K and \lgg\ $= 4.0 \pm 0.5$, 
and for the secondary \teff\ $=7500 \pm 300$K and \lgg\ $= 4.0 \pm 0.5$.

HD~98088 was observed with Hipparcos, allowing us to determine accurate luminosities.  
Based on the new reduction by \citet{van_Leeuwen2007-Hipparcos_validation}, 
the system has a parallax of $7.72 \pm 0.4$ mas and a $V$ magnitude of $6.173 \pm 0.007$. 
The uncertainty in magnitude is the standard deviation of the Hipparcos measurements rather 
than the (substantially smaller) quoted formal uncertainty, and agrees with the 
$b$ and $y$ band photometric variability found by \citet{Catalano1994-Ap-phot-var}.
The bolometric correction for Ap stars from \citet{Landstreet2007-surveyFORS1} 
was used, since we find both components of HD~98088 have strong peculiarities, 
producing $BC_{\rm A} = 0.08$ and $BC_{\rm B} = 0.09$ (we adopted $M_{\rm bol,\odot}=4.74$).  
For comparison, the bolometric correction of \citet{Balona1994-bolometric-corr} 
produces $BC_{\rm A} = 0.01$ and $BC_{\rm B} = 0.04$.  
Following the recommendation of \citet{Landstreet2007-surveyFORS1}, 
we adopted a conservative uncertainty on the bolometric correction of 0.1 in order 
to account for any potential systematic uncertainties.  
Using the luminosity ratio we calculated the bolometric magnitude difference between 
the primary and the total system ($m_{\rm A}-m_{\rm tot} = 0.265\pm0.053$), 
and the difference between the secondary and the total system ($m_{\rm B}-m_{\rm tot} =1.662\pm0.191$).  
The final luminosities we derive are shown in Table \ref{fundimental-param}.

\begin{figure*}
\centering
\includegraphics[width=4.5in]{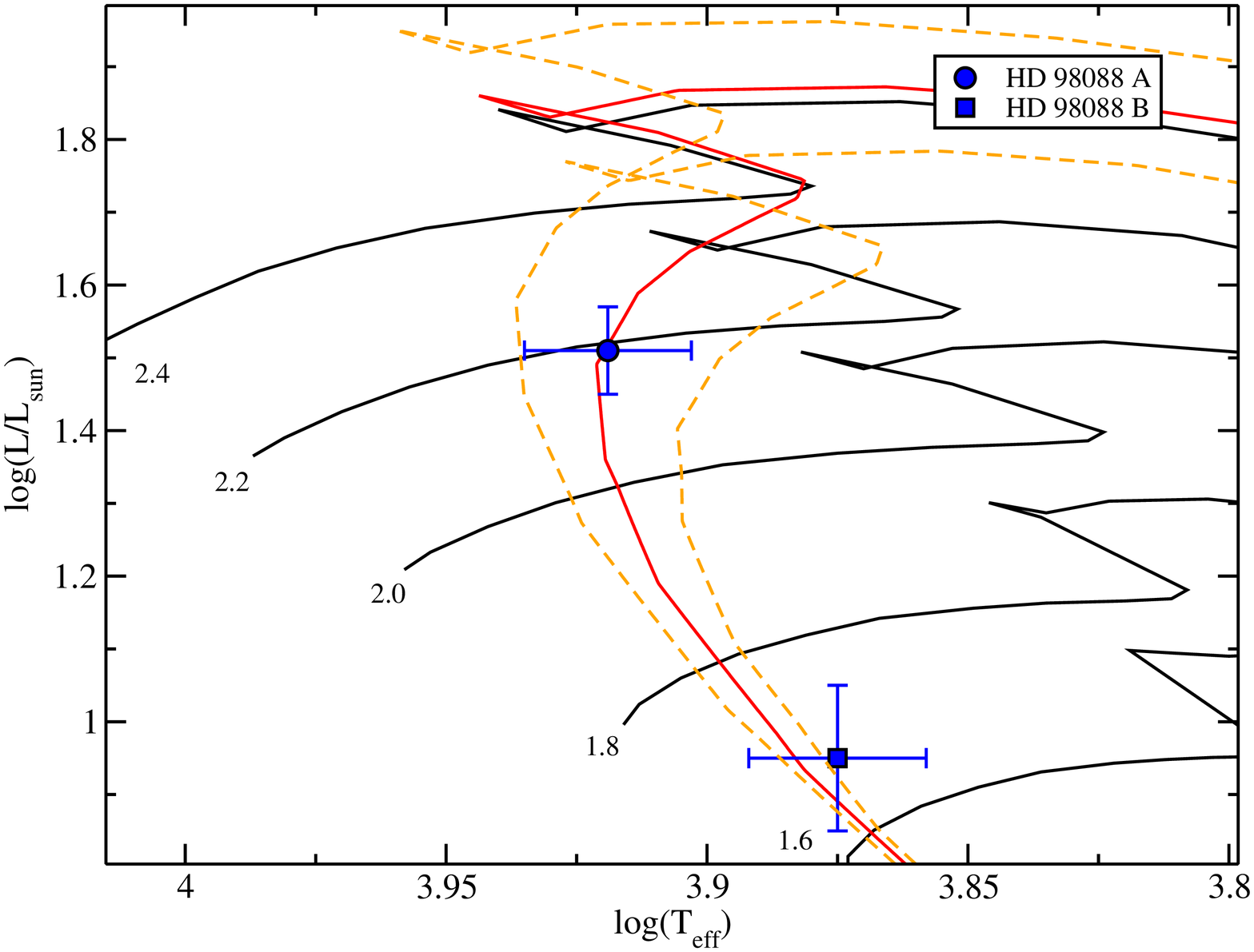}
\caption{The positions of HD~98088 A and B on the H-R diagram.  Evolutionary tracks (black) 
are labelled in units of solar masses, and 
isochrones correspond to the age of the system ($8.81 \pm 0.06$ log yr). 
Evolutionary tracks and isochrones are from \citet{Schaller1992-ms-evol} for 
standard mass loss and $Z=0.02$.  }
\label{hr-diagram}
\end{figure*}

With luminosities for both components, we can place the stars on the H-R diagram. 
By comparison with the evolutionary tracks of \citet{Schaller1992-ms-evol} computed 
with $Z=0.02$ and standard mass-loss, we determined masses for both components 
and an age for the system.  The H-R diagram positions of both stars are 
consistent with a single isochrone, suggesting that the stars are coeval.  
The age is determined from the H-R diagram position of the primary; 
since it is the more evolved star it allows us to determine a more precise age. 
The luminosities, masses, and age of the system are reported in 
Table \ref{fundimental-param}, and the H-R diagram of the system is presented 
in Fig.\ \ref{hr-diagram}.

\begin{table}
\centering
\caption{Summary of physical parameters for the HD 98088 system. }
\begin{tabular}{lcc}
\hline
                      & HD~98088~A      & HD~98088~B \\
\hline
$d$ (pc)              & \multicolumn{2}{c}{$129.5\pm6.7$} \\
$i_{\rm orb}$ ($\degr$) & \multicolumn{2}{c}{$69 \pm 2$}    \\
$\log\,{\rm age}$     & \multicolumn{2}{c}{$8.81\pm0.06$} \\
\teff (K)             & $8300\pm300$    & $7500\pm300$ \\
\lgg\ (cgs)           & $3.90\pm0.09$   & $4.16\pm0.12$  \\ 
${\log}\,{L/L_{\odot}}$ & $1.51\pm0.06$   & $0.95\pm0.10$  \\
$M$ (\Msun)           & $2.19\pm0.07$   & $1.67\pm0.08$  \\
$R$ (\Rsun)           & $2.76\pm0.28$   & $1.77\pm0.24$  \\
\hline
\end{tabular}
\label{fundimental-param}
\end{table}

A few additional useful parameters can be calculated, and a few consistency 
checks can be made between our H-R diagram parameters 
and our dynamic parameters from the radial velocity curve.  The mass ratio 
from the H-R diagram $M_{\rm A}/M_{\rm B} = 1.31\pm0.08$ agrees well with 
the more precise dynamic value $M_{\rm A}/M_{\rm B} = 1.374 \pm 0.008$.  
Combining the dynamic $M_{\rm A} \sin^3 i$ with the H-R diagram $M_{\rm A}$ 
we derive the inclination of the orbital axis with respect to the line of sight 
$i_{\rm orb} = 69 \pm 2 \degr$.  Using the secondary's mass we find the consistent 
value $i_{\rm orb} = 67 \pm 2 \degr$.  
Then the semi-major axis of the primary's orbit can be calculated, using 
the dynamic $a_{\rm A} \sin i$, yielding $a_{\rm A} = 9.03 \pm 0.10$ \Rsun. 
Similarly, for the secondary $a_{\rm B} = 12.40 \pm 0.15$ \Rsun.
We also calculate the inclination of the rotational axis of the primary, 
using the rotational period from the magnetic analysis and \vs\ from 
the abundance analysis, and find $i_{\rm rot,A} = 68^{+22}_{-17} \degr$.  
While this is rather uncertain, it is consistent with the 
orbital inclination, which one would expect for a tidally locked system. 
We can also calculate \lgg\ based on the star's H-R diagram masses and radii, 
finding \lgg~$= 3.90\pm0.09$ for the primary and \lgg~$= 4.16\pm0.12$ 
for the secondary.  These are consistent with the values from the SED, 
but significantly more precise.  

We can also compare our values to the detailed work of \citet{Carrier2002-binarity_in_Ap},  
and we find our physical parameters are fully consistent with theirs.  
Our orbital periods and inclinations are consistent, as are our \teff, \lgg, 
luminosities, masses and radii.

\section{Chemical Abundances}
\label{Chemical Abundances}

A detailed abundance analysis was performed for both HD~98088~A and B, 
using the disentangled spectra.  Two independent procedures were adopted, 
and independent analyses were performed by two of us. 
One analysis (performed by K.L.) employed the LTE spectrum synthesis code 
{\sc Synth3} \citep{Kochukhov2007-synth3} coupled to the IDL visualisation 
script {\sc BinMag}.  The other analysis (performed by C.P.F.) 
employed a modified version of the {\sc Zeeman} spectrum 
synthesis code \citep{Landstreet1988-Zeeman1,Wade2001-zeeman2_etc} 
which solves the polarized radiative transfer equations, again assuming LTE. 
Optimizations to the code for negligible magnetic fields were included, 
and a Levenberg-Marquardt $\chi^2$ minimisation routine was used to aid in 
fitting the observed spectrum \citep{Folsom2012-HAeBe-abundances}.  

In both cases, atomic data was extracted from the Vienna Atomic Line Database (VALD) 
\citep{Kupka1999-VALD}, using an `extract stellar' request, with temperatures 
tailored to the components of HD~98088. 
The model atmospheres used in the abundance analysis were calculated using 
the {\sc LLModels} code \citep{Shulyak2004-llmodels}, which produces LTE model 
atmospheres with detailed line blanketing.  
Prior to abundance analysis, the average disentangled spectra of the primary 
and secondary were corrected for the continuum dilution by the other component 
\citep[see e.g.][]{Folsom2008-hd72106} using the theoretical, wavelength 
dependent light ratio predicted by the adopted LLModels atmospheres.

The {\sc Synth3} analysis involved directly fitting synthetic spectra to 
the deconvolved spectra of both components of HD~98088. 
The fitting was performed independently on a number of small spectral windows. 
For the primary, 20 spectral windows were used with typical lengths of 100 \AA\ 
(ranging from 50 to 150 \AA\ long), running from 4500 to 6500~\AA, 
excluding Balmer lines. 
For the secondary, 13 windows were used typically with 150 \AA\ lengths, 
again running from 4500 to 6500~\AA\ excluding Balmer lines.
The fitting was performed by manually adjusting input chemical abundances, \vs, 
and microturbulence, then evaluating the fit of the synthetic spectrum 
to the observation with the {\sc BinMag} visualisation tool.  
The best fit values from the individual windows were averaged to produce 
the final best values, and the uncertainties on the final best values were 
taken to be the standard deviation of the values from individual windows.  
For elements with abundances derived only in three or fewer windows, 
the uncertainty was taken to include the line-to-line scatter as well as 
the range of abundances produced by varying \teff\ and \lgg\ by $1\sigma$.  
The final best fit values are reported in Table~\ref{abun-tab}.

\begin{figure*}
\centering
\includegraphics[width=4.5in]{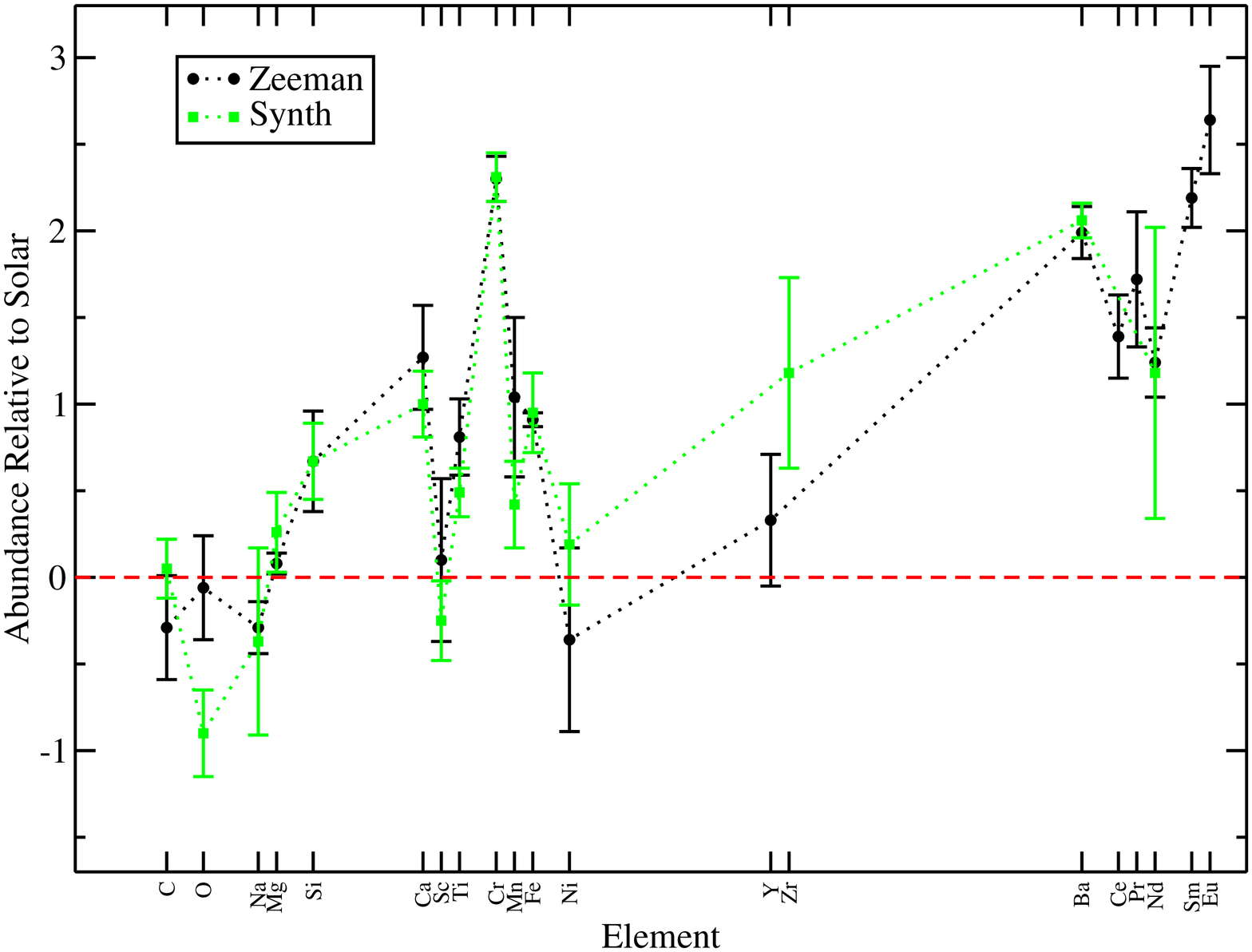}
\includegraphics[width=4.5in]{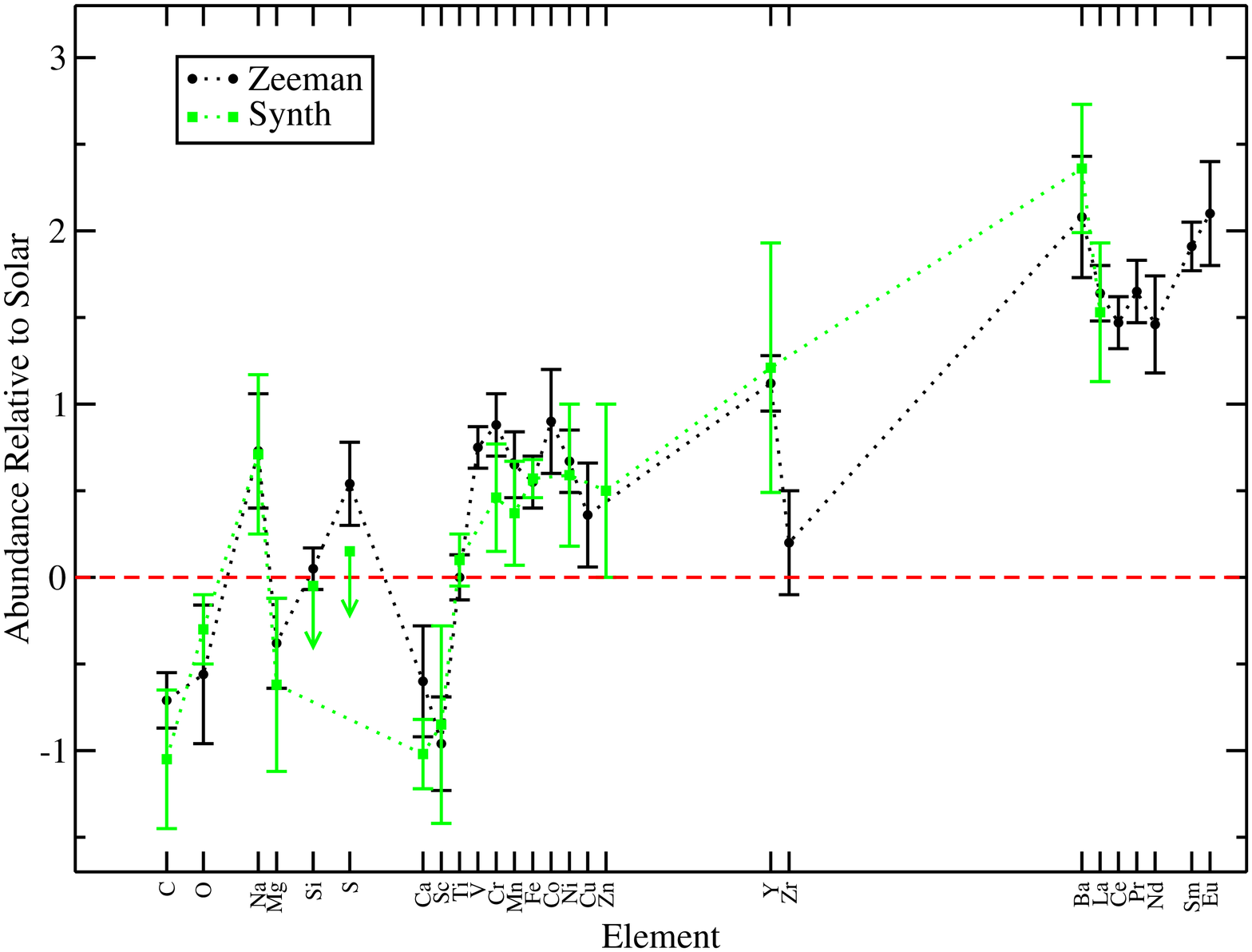}
\caption{Abundances relative to solar for HD~98088~A (top frame) 
and HD~98088~B (bottom frame), averaged over all spectral windows modelled.  
The results from both the analysis with {\sc Zeeman} (circles, black) 
and {\sc Synth3} (squares, green) are presented.  
Solar abundances are taken from \citet{Asplund2009-solar-abun}.
Arrows represent upper limits only.   }
\label{abunplot}
\end{figure*}

The {\sc Zeeman} analysis proceeded by fitting the spectra of both stars 
independently for chemical abundances, as well as \vs\, and microturbulence.  
Fitting was performed for 6 independent spectral regions, between 200 and 400~\AA\ long,
covering a spectral range from 4900 to 6500 \AA.  The final best fit values adopted are averages 
of the results over all 6 windows, with the adopted uncertainty as the standard 
deviation of those results.  For elements with less than four lines available, 
the the uncertainty was estimated by eye instead, including potential normalisation 
errors, blended lines, and the scatter between lines.  
The final averaged best fit values and uncertainties are presented in 
Table~\ref{abun-tab}.  The best fit chemical abundances are plotted, 
relative to the solar values of \citet{Asplund2009-solar-abun}, 
in Fig.~\ref{abunplot} for both components of HD~98088.

\begin{table*}
\centering
\caption{Summary of abundances, \vs, and microturbulence ($\xi$) for HD~98088~A and B, 
for both the analysis with {\sc Synth3} and {\sc Zeeman}.  
Chemical abundance are in units of $\log(N_{X}/N_{tot})$.
For the {\sc Synth3} results, abundances fit manually with lines from three or fewer 
spectral regions are marked with an asterisk (*).  
For the {\sc Zeeman} results, abundances fit with three or less useful lines 
are marked with an double-asterisk (**). 
Solar abundances are from \citet{Asplund2009-solar-abun}.  }
\begin{tabular}{cccccc}
\hline \hline \noalign{\smallskip}
        & HD 98088 A       & HD 98088 B       & HD 98088 A       & HD 98088 B       &  Solar \\
        & {\sc Synth3}     & {\sc Synth3}     & {\sc Zeeman}     & {\sc Zeeman}     &        \\   
\noalign{\smallskip} \hline \noalign{\smallskip}                                             
\vs\ (\kms) &$22.0 \pm 1.5$& $18.0 \pm 2.0$   & $24.4\pm0.8$     & $20.6\pm0.8$     &        \\
$\xi$ (\kms)&$ 1.0 \pm 0.5$& $ 1.5 \pm 0.5$   & $2.0$            & $ 1.9\pm0.2$     &        \\
\noalign{\smallskip} \hline \noalign{\smallskip}                                             
C       & -3.52 $\pm$ 0.17*& -4.62 $\pm$ 0.40*& -3.86$\pm$0.30** & $-4.28\pm0.16$   & -3.57 \\
O       & -4.21 $\pm$ 0.25*& -3.61 $\pm$ 0.20*& -3.38$\pm$0.30** & $-3.87\pm0.40$** & -3.31 \\ 
Na      & -6.13 $\pm$ 0.54*& -5.05 $\pm$ 0.46 & -6.05$\pm$0.15** & $-5.03\pm0.33$** & -5.76 \\
Mg      & -4.14 $\pm$ 0.23 & -5.02 $\pm$ 0.50*& -4.32$\pm$0.06   & $-4.78\pm0.26$   & -4.40 \\
Si      & -3.82 $\pm$ 0.22 & $\leq$-4.54      & -3.82$\pm$0.29   & $-4.44\pm0.12$   & -4.49 \\
S       &                  & $\leq$-4.73      &                  & $-4.34\pm0.24$   & -4.88 \\
Ca      & -4.66 $\pm$ 0.19 & -6.68 $\pm$ 0.20 & -4.39$\pm$0.30   & $-6.26\pm0.32$   & -5.66 \\
Sc      & -9.10 $\pm$ 0.23 & -9.70 $\pm$ 0.57 & -8.75$\pm$0.47   & $-9.81\pm0.27$   & -8.85 \\
Ti      & -6.56 $\pm$ 0.14 & -6.95 $\pm$ 0.15 & -6.24$\pm$0.22   & $-7.05\pm0.13$   & -7.05 \\
V       &                  &                  &                  & $-7.32\pm0.12$   & -8.07 \\
Cr      & -4.05 $\pm$ 0.14 & -5.90 $\pm$ 0.31 & -4.06$\pm$0.13   & $-5.48\pm0.18$   & -6.36 \\
Mn      & -6.15 $\pm$ 0.25*& -6.20 $\pm$ 0.30*& -5.53$\pm$0.46   & $-5.92\pm0.19$   & -6.57 \\
Fe      & -3.55 $\pm$ 0.23 & -3.93 $\pm$ 0.11 & -3.59$\pm$0.04   & $-3.95\pm0.15$   & -4.50 \\
Co      &                  &                  &                  & $-6.11\pm0.30$** & -7.01 \\
Ni      & -5.59 $\pm$ 0.35 & -5.19 $\pm$ 0.41 & -6.14$\pm$0.53   & $-5.11\pm0.18$   & -5.78 \\
Cu      &                  &                  &                  & $-7.45\pm0.30$** & -7.81 \\ 
Zn      &                  & -6.94 $\pm$ 0.50*&                  &                  & -7.44 \\
Y       &                  & -8.58 $\pm$ 0.72 & -9.46$\pm$0.38   & $-8.67\pm0.16$** & -9.79 \\
Zr      & -8.24 $\pm$ 0.55*&                  &                  & $-9.22\pm0.30$   & -9.42 \\
Ba      & -7.76 $\pm$ 0.10 & -7.46 $\pm$ 0.37 & -7.83$\pm$0.15   & $-7.74\pm0.35$** & -9.82 \\
La      &                  & -9.37 $\pm$ 0.40*&                  & $-9.26\pm0.16$   & -10.90\\
Ce      &                  &                  & -9.03$\pm$0.24   & $-8.95\pm0.15$   & -10.42\\ 
Pr      &                  &                  & -9.56$\pm$0.39   & $-9.63\pm0.18$   & -11.28\\
Nd      & -9.40 $\pm$ 0.84 &                  & -9.34$\pm$0.20   & $-9.12\pm0.28$   & -10.58\\
Sm      &                  &                  & -8.85$\pm$0.17   & $-9.13\pm0.14$   & -11.04\\
Eu      &                  &                  & -8.84$\pm$0.31** & $-9.38\pm0.30$** & -11.48\\ 
\noalign{\smallskip} \hline \hline
\end{tabular}
\label{abun-tab}
\end{table*}

\subsection{Abundances of HD 98088 A}

\begin{figure*}
\centering
\includegraphics[width=5.0in]{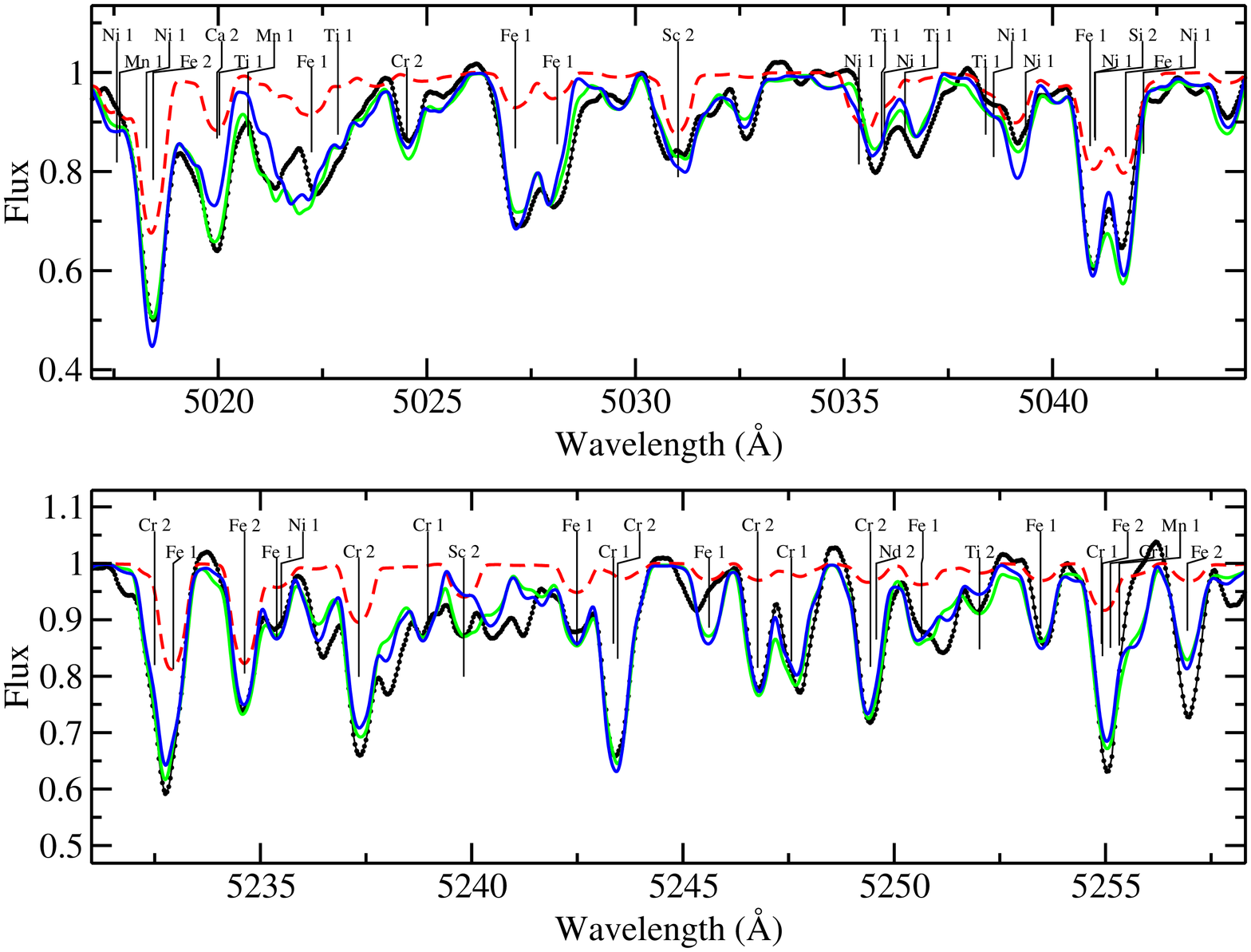}
\caption{Typical fits to the spectrum of HD~98088~A for both the {\sc Zeeman} 
and {\sc Synth3} results.  Points represent the observation, the lighter grey 
(green) line is the {\sc Zeeman}, and the darker grey (blue) line is the 
{\sc Synth3} best fit spectrum.  The dashed line is a synthetic spectrum with 
solar abundances. 
Elements which are major contributors to each line have been labelled.  }
\label{fit-primary}
\end{figure*}

The best fit synthetic spectra from both the {\sc Zeeman} and the {\sc Synth3} 
analyses generally reproduce the observed disentangled spectrum well.  
Samples of typical fits to the observation for both analyses of HD~98088~A are 
presented in Fig.~\ref{fit-primary}.  Small differences between both synthetic spectra 
and the observation do exist.  This is likely due to chemical stratification, 
which is common in Ap stars, or to blending with unknown rare-earth lines, 
which can also occur in Ap stars at these temperatures.   

HD~98088~A clearly shows typical Ap peculiarities, with strong overabundance 
of Si, Ca, Ti, Mn, Fe, and Ni, a dramatic overabundance of Cr, and very strong 
overabundances of Ba, Ce, Pr, Nd, Sm and Eu.  Ni and Sc have nearly 
solar abundances, as do C, O, Na, and Mg.  

In order to approximate the desaturation effect of the magnetic 
field of HD~98088~A in our spectra, a microturbulence of 2 \kms\ 
was assumed for the {\sc Zeeman} analysis and a value of 1 \kms\ 
was used for the {\sc Synth3} analysis.  While the difference in assumed 
microturbulence was a consequence of the independent analyses, it serves to illustrate 
the minor impact that small variations in microturbulence have on our results.  
There are a half-dozen careful analyses of sharp lined Ap stars published in the literature 
\citep[e.g.][from Ryabchikova private communication]{Ryabchikova1997-roAp3-gamma-Equ, Kochukhov2006-reg-slo-ver-inver},
all of which find microturbulence consistent with zero.  
Thus, while the typical microturbulence in an Ap star is unknown, it is likely small.  
The use of microturbulence for HD~98088~A 
is simply to approximate line broadening due to the magnetic field, 
thereby reducing the computation time for spectra dramatically.  
Replacing the microturbulence with a dipole magnetic field like that determined in 
Sect. \ref{magnetic-fields} would modify most of our abundances by 0.1-0.2 dex, 
which is generally within our uncertainties.  

The abundance of HD~98088~A derived from the two analyses agree very well.  
The abundances all agree within $1\sigma$, except for Ti and Mn 
at $1.2\sigma$ and O at $2\sigma$.  Both O abundances are based 
on one weak, heavily blended triplet of lines (6156-6158 \AA), thus an additional 
systematic effect may be present, or the uncertainties may be underestimated.  
The \vs\ values agree at $1.4\sigma$.  No systematic differences 
between the results are found.  

An attempt was made to determine a spectroscopic \teff\ and \lgg\ by fitting metal 
lines of varying excitation potentials and ionisation states using the {\sc Zeeman} 
program and the method described by \citet{Folsom2012-HAeBe-abundances}.  
This attempt was unsuccessful.  While we found \lgg~$=4.1 \pm 0.2$ which appears 
to be accurate, we also found \teff~$=9000 \pm 200$ K which is clearly 
inconsistent with the observed SED.  A likely cause of this discrepancy is chemical 
stratification in HD~98088~A, which could cause lines formed deeper in the stellar 
atmosphere to be anomalously strong, thereby producing an incorrect \teff\ estimate 
when metallic lines are used.  

The presence of chemical stratification in HD~98088~A was investigated using 
the {\sc Zeeman} spectrum synthesis program.  While stratification is not the 
focus of this study, these results could prove instructive for future studies, 
and they help explain the initial inaccurate spectroscopic \teff.  
Stratification was modelled with a simple step function, consisting of 
one abundance deeper in the atmosphere, another abundance higher in the atmosphere, 
and an optical depth for the transition.  In this pilot study Fe stratification 
was investigated, while other elements were assumed to have homogeneous abundances.  
The optical depth scale used was the optical depth (due to continuous opacity) 
at the middle of the wavelength region modelled. 
The Fe stratification parameters were added to the $\chi^2$ minimisation routine, 
and they as well as all other abundances and \vs\ were fit simultaneously, 
while the atmospheric parameters were held fixed at the values in Table \ref{fundimental-param}.  
The fitting was performed for the 4900-5500 \AA, 5500-5700 \AA, 
and 6100-6500 \AA\ regions independently.  The resulting fits provide a 
clear improvement over a homogeneous model, both by visual inspection 
and by $\chi^2$.  Averaging over the three windows, and taking the range of values 
as the uncertainty, the Fe abundance higher in the atmosphere (lower optical depth) 
is ${\rm Fe}_{upper} = -4.5 \pm 0.2$ dex, the Fe abundance deeper in the atmosphere 
is ${\rm Fe}_{lower} = -2.2 \pm 0.3$ dex, and the transition occurs at 
$\log \tau = -0.3 \pm 0.2$. 
While this represents strong Fe stratification, it is similar to what is seen 
in other Ap stars of similar \teff\ 
\citep[e.g.][]{Ryabchikova2002-stratification-gamma-Equ,Ryabchikova2005-strat-HD204411,Kochukhov2009-alphaCir-strat}.

We then attempted to determine a spectroscopic \teff\ simultaneously with Fe 
stratification (relying principally on the ionisation balance for \teff).  
This was done by using the $\chi^2$ minimisation routine to fit \teff, 
Fe stratification, \vs, and other chemical abundances simultaneously, again 
using the 4900-5500 \AA, 5500-5700 \AA, and 6100-6500 \AA\ spectral regions.  
We find \teff~$= 8500 \pm 300$ K, ${\rm Fe}_{upper} = -4.0 \pm 0.2 $ dex, 
${\rm Fe}_{lower} = -2.2 \pm 0.4$ dex, and $\log \tau = -0.5 \pm 0.3$.  
This \teff\ is consistent with the SED of HD~98088.  The stratification result is
is also consistent with the previous values, though the small difference in 
${\rm Fe}_{upper}$ suggests it may be sensitive to the exact \teff\ used.  
As a note of caution, since only the stratification of Fe was modelled here, 
stratification of other elements could be influencing the derived \teff\ 
potentially making it slightly inaccurate.  However, this model 
does succeed in reconciling the spectroscopic and spectrophotometric observations.  

\subsection{Abundances of HD 98088 B}

\begin{figure*}
\centering
\includegraphics[width=5.0in]{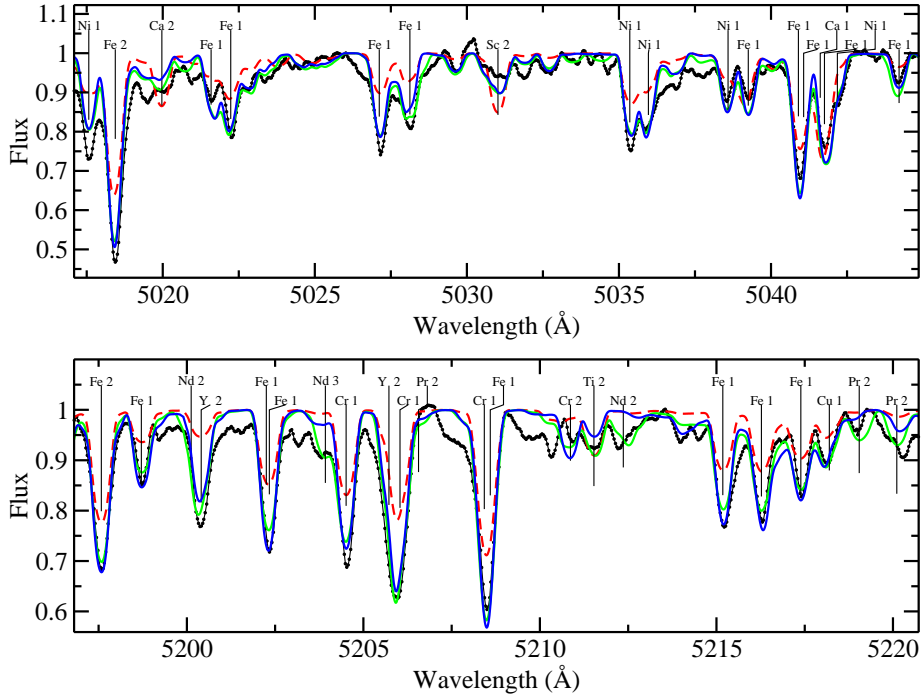}
\caption{Typical fits to the spectrum of HD~98088~B for both the {\sc Zeeman} 
and {\sc Synth3} results.  Points represent the observation, the lighter grey 
(green) line is the {\sc Zeeman}, and the darker grey (blue) line is the 
{\sc Synth3} best fit spectrum.  The dashed line is a synthetic spectrum with 
solar abundances. 
Elements which are major contributors to each line have been labelled.   }
\label{fit-secondary}
\end{figure*}

We achieve good fits to the observed spectrum of HD~98088~B for both the 
{\sc Synth3} and {\sc Zeeman} analyses.  Typical fits from both analyses 
of HD~98088~B are presented in Fig.~\ref{fit-secondary}. 
Both sets of best fit abundances for HD~98088~B are reported in 
Table~\ref{abun-tab}, and plotted relative to solar in Fig.~\ref{abunplot}.

The abundances we find reveal that HD~98088~B is a clear Am star.  
We find strong overabundances of Cr, Mn, Fe, Co, Ni, and Y, and very strong 
overabundances of Ba, La, Cr, Pr, Nd, Sm and Eu.  Sc and Ca are strongly 
underabundant, C appears to be underabundant, and O and Mg are marginally underabundant.  
This pattern of abundances, with overabundant Fe-peak elements and rare-earths, 
and underabundant Ca and Sc is characteristic of Am stars.  
We also find a small microturbulence of $\approx 2$ \kms.

In the {\sc Zeeman} analysis, microturbulence was determined 
by including it as a free parameter in the $\chi^2$ analysis and fit 
simultaneously with chemical abundances.  
In the {\sc Synth3} analysis, microturbulence was determined by fitting 
the synthetic spectrum to the observation by eye, together with chemical abundances.  
Microturbulence is often elevated in Am stars 
\citep[typically 2 to 4 \kms, e.g.][]{Landstreet2009-vmic2},  
and thus can have a significant impact on the abundance analysis of these stars. 

The results for the two abundance analyses agree very well.  
The chemical abundances, \vs, and microturbulence all agree within $1\sigma$, 
except for Ca and Cr which agree at $1.1\sigma$ and $1.2\sigma$ respectively.  
This confirms the accuracy of our methodology.  
Our uncertainties, if anything, may be slightly overestimated,
since statistically one would expect a few of our results 
to deviate by more than $1\sigma$.  

We attempted to determine a spectroscopic \teff\ for HD~98088~B using fits 
to metallic lines, similarly to HD~98088~A.  This was done using {\sc Zeeman} 
and independent fits for all 6 windows used for the abundance analysis.  
The produced a value of \teff~$= 8000 \pm 300$ K, which is consistent with 
the result from the SED at the $1.2 \sigma$ level.  
We prefer the SED value, since the spectroscopic \teff\ could be influenced 
by any small systematic errors in the disentangled spectrum, 
or by inaccurate or missing atomic data, particularly for the strongly 
overabundant rare-earth elements.  

This is the first report of chemical peculiarities in HD~98088~B, 
and a very rare case of a binary system with an Ap primary and an Am secondary.  
However, the appearance of Am peculiarities in HD~98088~B may reflect 
the generally higher incidence of Am peculiarities in low \vs\ stars 
\citep{Abt1995-Astar-vsini}, and in binary systems 
\citep[e.g.][]{Carquillat2007-survey-Am-SB}.  

\section{Magnetic fields of the components}
\label{magnetic-fields}

The magnetic fields in the photospheres of HD~98088~A and B 
were studied using the circular and linear polarisation within the LSD profiles, 
produced as a consequence of the longitudinal and transverse Zeeman effect. 

\begin{figure*}
\centering
\includegraphics[width=12cm,angle=-90]{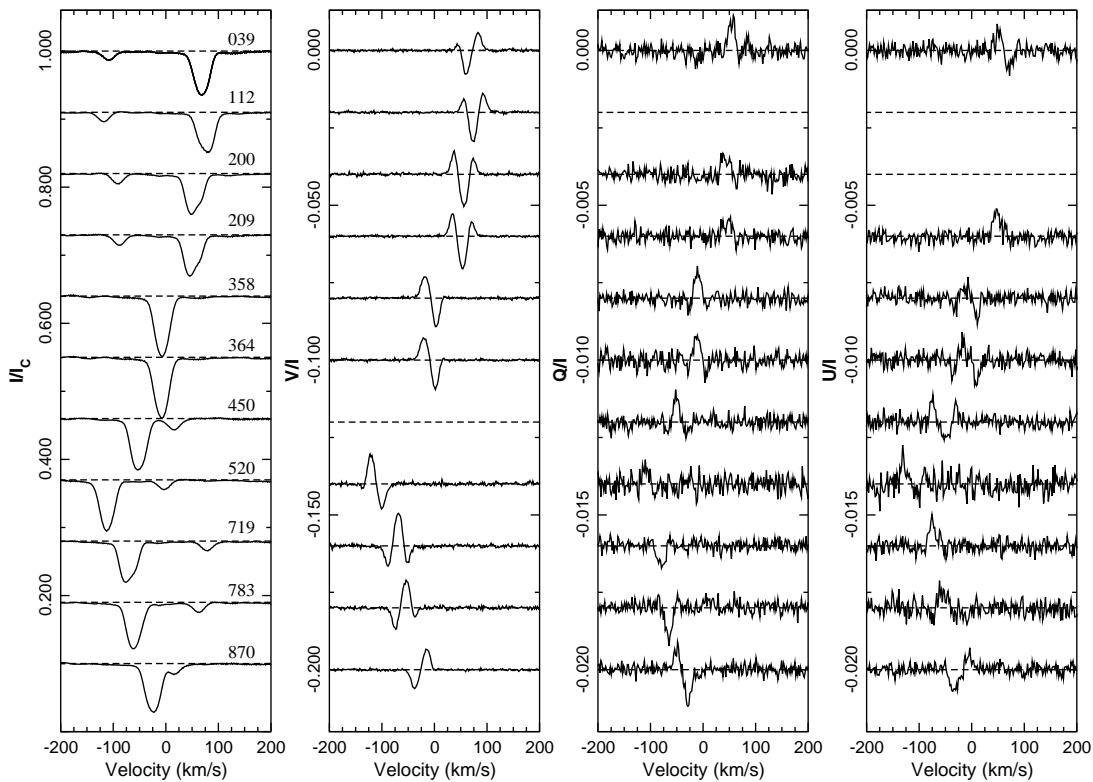}
\caption{LSD Stokes $IVQU$ profiles extracted from the observations of HD 98088. The 
weaker line of the secondary is visible in the leftmost frame. Polarisation 
signatures are observed at the position of the mean line of the primary, and follow 
the orbital motion of that line. No polarisation is observed at the position of the 
mean line of the secondary.  Phases ($\times 1000$) are labelled in the leftmost frame. }
\label{lsdprofs}
\end{figure*}

Fig.~\ref{lsdprofs} illustrates the extracted Stokes $I$, $V$, $Q$ and $U$ LSD profiles, 
ordered according to orbital phase. In the leftmost frame, the mean Stokes $I$ profiles 
of both the primary and secondary stars are visible, and exhibit systematic 
displacement as a consequence of their orbital motion. The relative weakness of the 
secondary's mean line is notable. In the remaining 3 panels, it is clear that Stokes 
$V$, $Q$ and $U$ Zeeman signatures are detected only at the position of the mean 
line of the primary star, indicating (at least at first glance) that the secondary 
is non-magnetic. Indeed, evaluating the detection probability according to the 
criteria introduced by \citet{Donati1997-major} indicates that significant Stokes 
$V$ signal is detected in the mean line of the primary star at all phases, whereas 
no significant signal is ever detected in the mean line of the secondary. Finally, 
we point out the excellent phasing of the LSD profiles according to the orbital 
period. Although the data span more than 180 orbits of the system, there is 
excellent fidelity of the detailed profile shapes and positions inferred to correspond to similar 
phases, and acquired both close together or far apart in time (e.g. phases 
0.200/0.209 and 0.358/364). This is the most visually striking evidence for precise 
synchronization of the orbital period and the rotational period of the magnetic 
primary star. 

To quantify the magnetic properties of the two stars, we measured the longitudinal 
magnetic field \bz\ of both components from the LSD profiles using the first-moment 
method of \citet{Rees1979-magnetic-cog}. We integrated the $I/I_{\rm c}$ and $V/I_{\rm c}$ 
profiles about their centres-of-gravity $v_{\rm 0}$ in velocity $v$, in the 
manner implemented by \citet{Donati1997-major} and corrected by 
\citet{Wade2000-highPrecision-correctBz}: 
\begin{equation}
B_{\ell}=-2.14\times 10^{11}\ \frac{{\displaystyle \int (v-v_{\rm 0}) V(v)\ dv}}{\displaystyle {\lambda z c\ \int [1-I(v)]\ dv}}.
\label{bz-equation}
\end{equation}
In Eq.~\ref{bz-equation} $V(v)$ and $I(v)$ are the $V/I_{\rm c}$ and $I/I_{\rm c}$ LSD profiles,
respectively. The wavelength $\lambda$ is expressed in nm and the longitudinal field
\bz\ is in gauss. The wavelength and Land\'e factor $z$ correspond to the mean
values employed in computing the LSD profiles, reported in Sect.~\ref{observations}.

For the primary, the velocity integration limits for the evaluation of Eq.~\ref{bz-equation} 
were selected manually to include the entire observed Stokes $V$ profile.  In the case of 
the secondary, the limits were chosen to include the entire Stokes $I$ profile.  
The centre-of-gravity for each component was also calculated using these integration limits.  
Table \ref{magnetic_table} summarises the longitudinal magnetic field measurements for each 
date for which Stokes $V$ observations were obtained. At phases at which the profiles 
of the primary and secondary star were blended, we measured only the longitudinal 
field of the primary star by assuming that the secondary made no contribution to the 
Stokes $V$ profile, and adjusting the denominator of Eq.~\ref{bz-equation} by subtracting the 
average equivalent width of the secondary's LSD profile. Standard uncertainty 
propagation was used to determine the $1{\sigma}$ error bars on the longitudinal 
magnetic field measurements.  Uncertainties range from $15$ to $35$ G for the 
primary and from $78$ to $280$ G for the secondary component.  A significant 
longitudinal magnetic field is detected in the primary component at all phases.  
On the other hand, no significant magnetic field is ever detected in the secondary 
component.  

We point out that the value of the longitudinal field of a component of an 
SB2 system inferred using Eq. \ref{bz-equation} is unaffected by the continuum 
of the companion. This is because it is the continuum normalised 
Stokes $I$ and $V$ parameters ($I/I_{\rm c}$ and $V/I_{\rm c}$, 
respectively) that appear in the denominator and numerator of Eq.~\ref{bz-equation}. 
Nevertheless, the inferred uncertainty, derived from photon-noise error bars 
propagated through the LSD procedure, is larger than that for the single-star case. 
  
From these basic magnetic measurements, we conclude that the primary component of 
HD~98088 hosts a strong magnetic field with a peak disc-averaged longitudinal 
component of about 1100 G, whereas the secondary shows no detectable magnetic field. 

\subsection{Rotational period and magnetic geometry of the primary star}

The Lomb-Scargle periodogram of the longitudinal field measurements of the primary 
star displays a clear excess of power between 5.6 and 6.0~d, with many possible 
solutions in this range. The peak closest to the orbital period corresponds to a 
variability period of 5.90451~d, but with rather large uncertainties 
($\sim \pm0.0015$~d). If we combine our new longitudinal field measurements with 
those reported by \citet{Babcock1958-magnetic-catalog}, this yields a period 
of $5.90510\pm 0.00003$~d. This rotational period is in excellent agreement 
with the orbital period ($5.9051102\pm 0.0000023$; Table \ref{orbit-fit}). 
Given this correspondence and the better precision of the orbital period, 
hereinafter we assume perfect synchronization of the system 
and adopt the orbital period as the rotational period of both the primary 
and secondary stars. 

\begin{figure}
\centering
\includegraphics[width=6.5cm,angle=-90]{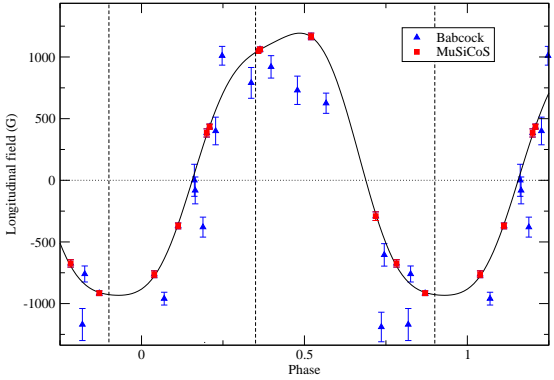}
\caption{Phased longitudinal magnetic field measurements of HD 98088~A. Filled 
squares\ - MuSiCoS measurements. Open triangles\ - Measurements reported by 
\citet[][]{Babcock1958-magnetic-catalog}. The vertical dashed lines indicate the 
approximate quadrature phases of the radial velocities. }
\label{mag-fld-plot}
\end{figure}

The phased longitudinal field measurements are shown along with those of 
\citet{Babcock1958-magnetic-catalog} in Fig.~\ref{mag-fld-plot}. The significant 
improvement in precision of the MuSiCoS data relative to Babcock's measurements is 
evident. Although the historical measurements are noisy, it appears that the new 
\bz\ curve may be shifted to slightly ($\sim 100-150$~G) higher values than the 
previous measurements. If real, we attribute this shift to differences in the 
modelling and analysis methods employed.  Such a difference is not surprising, 
since the older measurements were based on (somewhat non-linear) photographic data, 
with line centres estimated by eye.  Babcock's observations were made  
blueward of H$\beta$, while our observations are predominately to the red of that line, 
thus the two datasets are affected by limb darkening somewhat differently, 
which could also partly explain the discrepancy.  
Our new measurements show clear and systematic departures from a pure sinusoidal variation. 
Whereas first-order and second-order harmonic expansions fit phase variation of the 
new data rather poorly (reduced $\chi^2$ of 16.1 and 18.6, respectively), a third-order 
expansion achieves an excellent fit (reduced $\chi^2$ of 0.6). While these 
departures from a sinusoidal variation may diagnose non-dipolar characteristics of 
the stellar magnetic field, it is more likely that they reflect the non-uniform 
distributions of chemical elements whose lines were used to measure the field. 

\begin{figure}
\centering
\includegraphics[width=4.2cm,angle=-90]{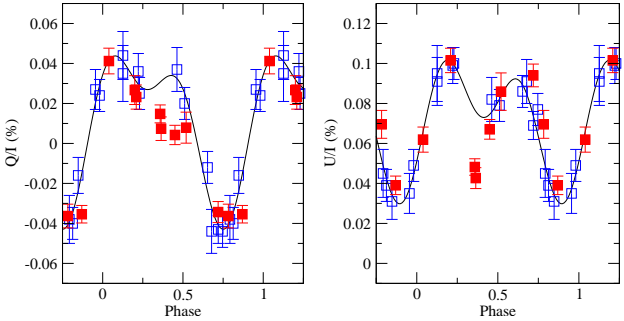}
\caption{Net linear polarisation of HD 98088~A, compared to the broadband linear 
polarization measurements of \citet{Leroy1995-broadband-pol}. As discussed by e.g.\ 
\citet{Silvester2012-data-paper}, the MuSiCoS measurements have been arbitrarily 
shifted and scaled to reproduce as closely as possible the broadband variations. }
\label{net-linear-plot}
\end{figure}

To determine the magnetic geometry of the primary, we took advantage of the Stokes 
$Q$ and $U$ LSD profiles to measure the net linear polarization (NLP), integrated 
across the LSD profiles \citep{Wade2000-highPrecision-correctBz,Silvester2012-data-paper}. 
These data are reported in Table~\ref{magnetic_table} and are compared to the 
broadband linear polarization (BBLP) measurements of HD~98088 reported by 
\citet{Leroy1995-broadband-pol}. in Fig.~\ref{net-linear-plot}. As described by e.g.\ 
\citet{Silvester2012-data-paper}, while NLP and BBLP result from the same physical 
mechanism, the methods of measurements are sufficiently different that their 
intercomparison requires that one data set be arbitrarily shifted and scaled 
relative to the other. The NLP values reported in Table~\ref{magnetic_table} and 
illustrated in Fig.~\ref{net-linear-plot} include this scaling and shifting. The 
BBLP and NLP phase variations are generally compatible, although some differences in 
detail are certainly present. These results are consistent with the general results 
reported by \citet{Wade2000-highPrecision-correctBz} and \citet{Silvester2012-data-paper}. 

In combination with the longitudinal field, the NLP/BBLP measurements provide an 
unambiguous estimate of the rotational axis inclination $i$, magnetic obliquity $\beta$ 
and polar field strength $B_{\rm d}$, under the assumption of a pure dipolar surface 
magnetic field configuration \citep{Leroy1995-broadband-pol}. A first-order harmonic fit 
to Babcock's data yields $B_{\ell,{\rm max}}=1050\pm 175$~G and $B_{\ell,{\rm min}}=-1280\pm 175$~G, 
giving $r=B_{\ell,{\rm max}}/B_{\ell,{\rm min}}=-0.82\pm 0.22$. An analogous fit to 
the MuSiCoS measurements yields $B_{\ell,{\rm max}}=1185\pm 65$~G and 
$B_{\ell,{\rm min}}=-1000\pm 65$~G, for $r=-0.84\pm 0.09$. The broadband linear polarization 
measurements of Leroy et al. give $s=-0.065\pm 0.13$, where $s$ is a (unitless) 
parameter describing the harmonic content of the NLP/BBLP phase variation, defined 
according to \citet{Leroy1995-broadband-pol}. For the NLP variation, this value is 
$+0.005\pm 0.20$. The combined values of $r$ and $s$ yield unique values for the 
angles $i$ and $\beta$. 

Using this method \citet{Leroy1996-broadband-pol-modelling} reported 
$i=85\pm 5\degr$ and $\beta=80\pm 5\degr$ for HD 98088~A. 
For the combination of BBLP + Babcock, we find 
$i=75\pm 15\degr$ and $\beta=73\degr\pm 3\degr$. For BBLP + MuSiCoS \bz, we find 
$i=77\pm 10\degr$ and $\beta=73\degr\pm 3\degr$. Finally, for NLP + MuSiCoS \bz, 
we find $i=77\pm 10\degr$ and $\beta=76\degr\pm 5\degr$. All of these values are 
in good mutual agreement, indicating that the systematic differences between the 
various data sets are not a major source of error. We also note that the derived 
values of the inclination $i$ are formally consistent with the orbital inclination 
determined from the masses ($68\pm 3\degr$). 

If we adopt the orbital inclination as the inclination of the rotation axis of the 
primary, along with $\beta=75\pm 5\degr$, we derive the surface polar strength of 
the primary's magnetic dipole to be $3850\pm 450$~G using the MuSiCoS measurements.

\subsection{Upper limit on magnetic field of the secondary star}

No magnetic signature is detected in the mean line of the secondary, nor is the 
longitudinal magnetic field detected with significance. However, we can estimate an 
upper limit on the dipole component of the secondary. Assuming the secondary's 
rotation is synchronized with the orbit, i.e. with rotational period equal to 
$P_{\rm orb}$ and rotation axis inclination $i=68\degr$, the median longitudinal 
field error bar of 160~G translates into a $3\sigma$ upper limit on the surface 
dipole of roughly 1550~G.

\begin{table*}
\centering
\caption{Magnetic measurements obtained from MuSiCoS spectra.  All phases 
are computed from the orbital ephemeris. HJD is from the Stokes V observation.  
Typical range of phase corresponding to a single Stokes $VQU$ sequence 
is approximately $0.01\times P_{\rm orb}$.}
\begin{tabular}{cccrrrr}
\hline
Date & ${\bar{\rm HJD}}$ & Orbital & \multicolumn{2}{c}{$B_\ell$ (G)}    & \multicolumn{2}{c}{NLP}\\
\    & (2,450,000+)      & Phase   &  Primary     & Secondary           & $Q/I$ (\%) & $U/I$ (\%)\\ 
\hline
 18 Jan 99 & 1197.617 &   0.358 &   $ 1048\pm  15$&                     & $ 0.015 \pm    0.004$ & $  0.048 \pm   0.004$ \\
 23 Jan 99 & 1202.590 &   0.200 &   $ 384\pm   33$&   $   51\pm    166$ & $ 0.027 \pm    0.007$ &                       \\
 24 Jan 99 & 1203.555 &   0.364 &   $ 1062\pm  19$&                     & $ 0.007 \pm    0.005$ & $  0.043 \pm   0.005$ \\
 05 Mar 00 & 1609.524 &   0.112 &   $-370\pm   24$&    $-210\pm    135$ &  \\
 07 Mar 00 & 1611.519 &   0.450 &                 &                     & $  0.004 \pm   0.005$ & $  0.067 \pm   0.005$ \\
 05 Dec 01 & 2249.685 &   0.520 &   $1167\pm   27$&    $-195\pm    193$ & $ 0.008 \pm    0.008$ & $  0.086 \pm   0.009$ \\
 07 Dec 01 & 2251.751 &   0.870 &   $-916\pm   19$&   $53\pm 78$        & $ -0.035 \pm   0.004$ & $  0.039 \pm   0.005$ \\ 
 08 Dec 01 & 2252.748 &   0.039 &   $-762\pm   28$&   $  -92\pm    133$ & $ 0.041 \pm    0.006$ & $  0.062 \pm   0.006$ \\ 
 09 Dec 01 & 2253.750 &   0.209 &   $ 435\pm   21$&   $  226\pm    112$ & $ 0.023 \pm    0.006$ & $  0.102 \pm   0.006$ \\ 
 12 Dec 01 & 2256.762 &   0.719 &   $-291\pm   35$&    $ 423\pm    280$ & $-0.034 \pm    0.005$ & $  0.094 \pm   0.006$ \\ 
 05 Jan 02 & 2280.759 &   0.783 &   $-674\pm   30$&    $-159\pm    156$ & $-0.036 \pm    0.006$ & $  0.070 \pm   0.007$ \\ 
\hline
\end{tabular}
\label{magnetic_table}
\end{table*}

\section{Discussion and Conclusions}

\begin{figure*}
\centering
\includegraphics[width=3.5in]{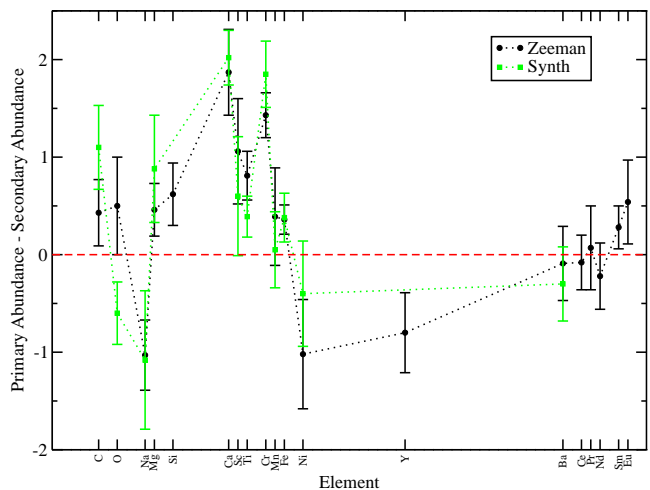}
\caption{Differential comparison of the chemical abundances found for the primary and secondary of HD~98088. 
Values from both the {\sc Synth3} (squares, green) analysis and the {\sc Zeeman} 
(circles, black) analysis are presented. }
\label{abunplot-comp}
\end{figure*}

\begin{figure}
\centering
\includegraphics[width=3.0in]{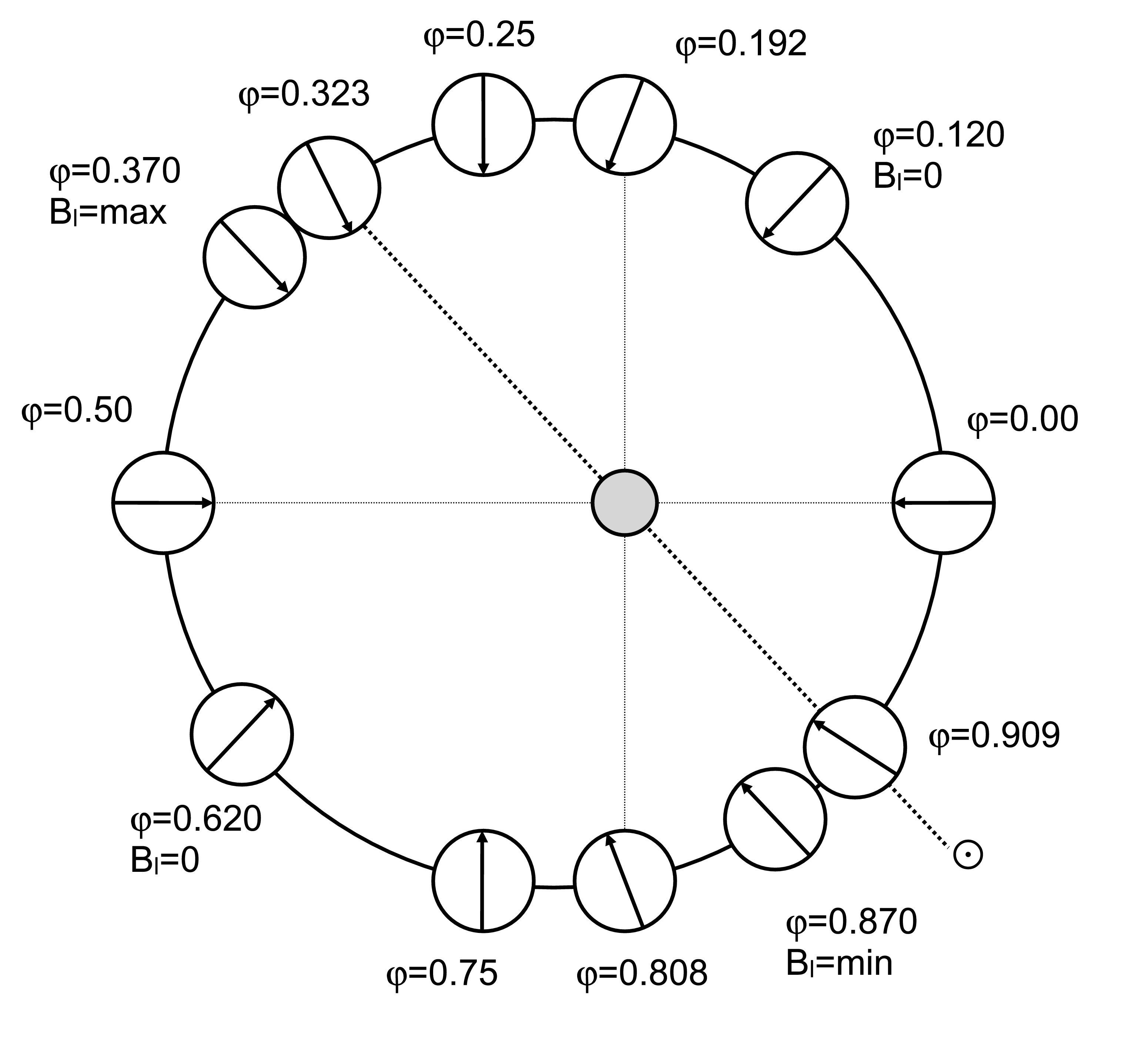}
\caption{Schematic of the orbit of HD~98088~A (circle with arrow) 
from the reference frame of HD~98088~B (filled circle).  
A dipole magnetic field for HD~98088~A is indicated with an arrow, 
assuming the dipole is aligned with the secondary at periastron. 
The dotted line represents the line towards the observer. 
The stellar radii and orbital separation are approximately to scale. 
Phases where the observed longitudinal magnetic field ($B_l$) is a maximum, minimum, 
and zero are shown, as are phases where the stars are aligned along the line of sight.  
Note that the phases where the radial velocities of the components are equal 
are shifted slightly from the phase where the stars are aligned along the line of sight 
(by +0.01 and -0.03 in phase), due to the eccentricity of the orbit.  
} 
\label{orbit-schematic}
\end{figure}

We find strong chemical peculiarities in HD~98088~A that are characteristic of a cool Ap star.  
In HD~98088~B we also find strong chemical peculiarities indicative of an Am star.
We also find evidence of strong Fe stratification in the atmosphere of the primary.  
The introduction of stratification was necessary to reconcile the spectroscopic 
and spectrophotometric temperatures of HD~98088~A.  
While Am stars are often found in close binary systems, HD~98088 represents 
one of very few close binary systems with an Ap component.  

Both components of HD 98088 have similar masses, \teff, \lgg, and \vs, 
yet the components display very different chemical peculiarities.  
Fig.~\ref{abunplot-comp} illustrates the difference in the derived abundances between the 
primary and the secondary.  The components of HD 98088 are likely coeval, but a qualitative  
difference is the presence of a strong magnetic field in the primary.  This suggests that 
the observed differences in chemical abundances are due to the impact of the magnetic field 
on atomic diffusion in the atmosphere of the primary \citep[e.g.][]{Michaud1981-diffusion_magneticApBp}.  

We find a strong, predominately dipolar, magnetic field in HD~98088~A, with a longitudinal 
strength varying from +1170 to -920 G.  Conversely, we find no magnetic field in 
the secondary with longitudinal field error bars down to 112 G, and a typical error bar of 160~G.  
We also find no signal from the secondary in linear polarisation, 
while the magnetic field of the primary is detected in linear polarisation at all phases.  
The predominately dipolar nature of the magnetic field in the primary is shown by the 
longitudinal field and net linear polarisation curves, and by the simple structure of 
the $V$, $Q$, and $U$ LSD profiles. 

It is not clear why the primary has a strong magnetic field and the secondary has 
no magnetic field when the two stars have similar masses and temperatures, 
and are apparently coeval with identical ages.  
Indeed, the origin of the magnetic fields in Ap stars 
is one of the enduring mysteries that these stars present. 

We derived improved orbital parameters for the HD~98088 system, finding a period of 
$5.9051102 \pm 0.0000023$ days.  We also derive a rotational period for 
HD~98088~A of $5.90510\pm 0.00003$ days, which is consistent with the orbital period.  
This provides strong evidence that the system is tidally locked, 
with the rotational and orbital periods synchronised.  
We derive an inclination of the orbital axis of the system of $i_{\rm orb} = 69 \pm 2 \degr$, 
which is consistent with the inclination of the rotational axis of the primary based on 
its rotational period, \vs, and radius, as well as the inclination derived from the 
linear polarisation under the assumption of a dipole magnetic field.  
However, the system has a modest but significant eccentricity of $0.1840 \pm 0.0025$, 
which is clearly visible in the radial velocity curve presented in Fig.~\ref{orbit-fit}. 
Thus while the system is synchronised, the orbits apparently have not had the time to circularise.  

\citet{Hut1981-binary-tidal-evolution} showed that, in binary systems where 
orbital angular momentum dominates over rotational angular momentum, tidal interactions 
cause the orbital and rotational axes to become parallel, and the orbital and rotational 
periods to become synchronised, much more quickly than they circularise the orbits of the system.  
Using Eq.~22 of \citet{Hut1981-binary-tidal-evolution} (and assuming an approximate 
radius of gyration of 0.2, following \citealt{stepien2000-Ap-rotation-momentum}), the ratio 
of orbital to rotational angular momentum is $\sim$150 for the primary, and $\sim$500 for the secondary.  
This suggests \citep[][Eq.~33]{Hut1981-binary-tidal-evolution} that circularisation of the orbits 
would be slower by a factor of $\sim$100 than the synchronisation of orbital and rotational periods.  
Thus it is certainly possible for HD~98088 to be rotationally synchronised but 
still have elliptical orbits.  However, we find an age for the system of $8.81\pm0.06$ $\log {\rm yr}$, 
which suggests that tidal interactions would have to be very inefficient for the system 
to still have some ellipticity.  

Modelling the magnetic field of HD~98088~A as a dipole, we find a dipole strength 
of  $3850\pm 450$~G and an obliquity angle of $\beta=75\pm 5\degr$. 
The longitudinal magnetic field reaches an extremum at very near the quadrature 
orbital phases (where the radial velocities of both components are near zero, 
see Figs.~\ref{orbit-fit} and \ref{mag-fld-plot}).  
Thus the magnetic dipole is close to being aligned along the axis between the two stars, 
with the positive magnetic pole always pointing near the secondary, 
as illustrated in Fig.~\ref{orbit-schematic}. 
This analysis does not account for the influence of abundance spots on the \bz\ curve, 
and if this effect were accounted for (with a self consistent modelling procedure, 
such as magnetic Doppler imaging) the alignment could be exact.  
This is likely significant, though the interpretation of this observation is not immediately obvious.  
It could be that the tidal interaction between the components somehow influenced 
the stable magnetic field geometry of the primary.  
For example if the magnetic field had a fossil origin as in the models of 
\citet{Braithwaite2006-stable-magnetic}, then tidal effects could affect 
the evolution of the magnetic field towards its stable configuration.  
However, magnetic fields in Ap stars appear to originate very early in 
a star's lifetime \citep[][in a few Myr]{Wade2005-HAeBe_Discovery}, 
and the orbital parameters may have changed since then.  
One could speculate that tidal interactions over the intervening years could 
have caused the field to evolve into its present configuration.  
Alternately, if the magnetic field perturbed the structure of the primary sufficiently 
to produce a small over-density at the magnetic poles, the system could settle into its 
current configuration through tidal effects.  

HD~98088 represents a very rare case of a magnetic Ap star in a tidally locked binary 
system with an Am star.  The magnetic field in the Ap star is predominately dipolar, 
with the positive pole always pointing towards the secondary.  
Further studies could help clarify the interpretation of this result.  
Magnetic Doppler imaging of HD~98088~A would be particularly valuable, as this would 
provide a more detailed model of the magnetic field, as well as a map of chemical 
inhomogeneities across the surface of the star.  These would be particularly interesting 
in the context of tidal interactions with the secondary.  
HD~98088 is an ideal target for the Binarity and Magnetic Interactions in Stars (BinaMIcS) project, 
and further observations and analysis are planned by the collaboration.

\section*{Acknowledgements} 
GAW is supported by an Natural Science and Engineering Research Council (NSERC Canada) Discovery Grant and a Department of National Defence (Canada) ARP grant.
OK is a Royal Swedish Academy of Sciences Research Fellow supported by grants from the Knut and Alice Wallenberg Foundation and the Swedish Research Council.

\bibliography{massivebib.bib}{}
\bibliographystyle{mn2e}

\label{lastpage}

\end{document}